\documentclass[envcountsect]{llncs}
\usepackage{url}
\usepackage{ifpdf}
 \ifpdf
\usepackage[pdftex]{graphicx}
 \else
\usepackage[dvips]{graphicx}
 \fi
\usepackage{amsmath}
\usepackage{amssymb}

\def\var#1{{\tt{#1}}}%%looks good!
\spnewtheorem{defi}{Definition}[section]{\bfseries}{\rmfamily}
\spnewtheorem{prop}{Proposition}[section]{\bfseries}{\rmfamily}
\spnewtheorem{fact}{Fact}[section]{\bfseries}{\rmfamily}
\spnewtheorem{coro}{Corollary}[prop]{\bfseries}{\rmfamily}

\begin{document}

\def\mytitle{
The Unreasonable Fundamental Incertitudes Behind Bitcoin Mining
 %\\
%{\small (extended version)}
%Is There A Better Method To Create Bitcoins?
%On the Hardness of Bitcoin Mining
%On Feasibility of the Fifth Generation of Bitcoin Miners
%Dark Side of The Bitcoin Market
%and of Cryptography Research
}
\title{\mytitle\\
\vskip-8pt
}

\author{
Nicolas T. Courtois$^{1}$,
Marek Grajek$^{2}$ \and
Rahul Naik$^{1}$
}

\institute{
$^{1}$
University College London, UK,
$^{2}$
Independent researcher and writer, Poland
}

\maketitle
%\centerline{\Large \mytitle}

%plan: position submission: 5th generation of Bitcoin miners
%later: covert generations before and after 5...
%cross-subsidies and new ways of doing business and subversive/conspiracy versions
%cite Adam Smith
%the new model for 5th generation of Bitcoin miners can be executed covertly

\vskip-8pt
\vskip-8pt
\vskip-8pt
\begin{abstract}
Bitcoin is a ``crypto currency'',
a decentralized %global virtual
electronic % and
payment scheme based on cryptography.
%%%%%invented
%%%%%and launched by crypto boffins in 2008/9,
%%%%%just after the current global financial crisis has started.
It implements a particular
type of peer-to-peer financial anarchy. % movement.
It has very recently gained excessive popularity and
attracted a lot of attention in the mainstream press and media.
%mostly from the practical, financial and legal point of view.
% and blogosphere,
%%both form the high tech and financial point of view
%cf.
%\cite{SummaryKrugman,NewYorkerBitcoinVeryDetailed,SteveForbesBitcoinsNotMoney,KrugmanAntiSocialNetwork}
%and many other.
Scientific research on this topic is less abundant.
%seems to be avoided and mocked!
A paper at Financial Cryptography 2012 conference explains that
Bitcoin is a system which
{\em uses no fancy cryptography, and is by no means perfect}
\cite{BitcoinFC12SecurityOverview}.
%nothing more to say, na fancy cryptography, not the best choices either
%%%and makes it very clear that modern cryptography
%%%allows to build much better electronic currency instruments.
%
%Bitcoin is a coin with two sides:
%a cryptographic, algorithmic and technical one,
%and a practical, social, economical and financial one.
%In this paper we look at both aspects of bitcoin:
%cryptographic, security engineering and financial innovation
%and social engineering.
%

Bitcoin depends on well-known cryptographic standards
such as SHA-256.
In this paper we revisit the cryptographic process
which allows one to
make money by producing new bitcoins.
We reformulate this problem
as a specific sort of Constrained Input Small Output
(CISO) hashing problem
and reduce the problem to a pure block cipher problem,
cf. Fig. \ref{BitcoinCISOProblem}.
We estimate the speed of this process
and we show that the amortized cost of this process is less than it seems
and it depends on a certain cryptographic constant
%which is estimated to be at most 1.86.
which is estimated to be at most 1.89.
These optimizations enable bitcoin miners to save %75,000 USD per day?
tens of millions of dollars per year in electricity bills.
%kncminer advertises sth like this, 30% saving, spotted on 15052013, dates to March/April or so
%this is based on advanced algos provided by orsoc.se
%evolution 150000=>175000=>200000 in April/July 2013 for Saturn advertised speed
%official on their web page: KNC miner is a joint venture between two strong companies, ORSoC AB and Kennemar & Cole AB
%
%We claim that a precise evaluation of this constant is likely to influence the market price of bitcoins in the future.
%More importantly the price will depend on the feasibility and supply of new very fast hardware miners.
%

Miners who set up mining operations face many economic incertitudes such as
%The incertitudes related to the adoption and
high volatility.
%of the bitcoin itself are widely known.
In this paper we point out that there are fundamental incertitudes
which depend very strongly on the bitcoin specification
and that this is specification
is NOT written in stone.  %said below, it is likely to change.
%One question is to see if bitcoin mining with ASICs
%is the last and final stage in the development of bitcoin.
The energy efficiency of bitcoin miners have already been improved
by a factor of about 10,000 since bitcoin have been invented,
and we claim that %take a bold view and predict
%that there will be a next generation of miners.
further improvements are inevitable.
Better technology is bound to be invented, would it be quantum bitcoin miners.
More importantly, the specification of the problem to solve is likely to change.
A major change in the Proof of Work function have been proposed in May 2013
at Bitcoin conference in San Diego by Dan Kaminsky \cite{KaminskyPredictsEndOfSHA256}.
However, any sort of change could be flatly rejected by the community
which have heavily invested in ASIC mining with the current technology.

Another question is the reward halving scheme in bitcoin.
The current bitcoin specification mandates a strong 4-year cyclic property.
This cycle is totally {\bf artificial} and is simply an artifact of the current specification.
We find this property totally unreasonable and harmful and
explain why and how it needs to be changed.

\end{abstract}

%ACM classifications E.3; D.4.6; %K.4.1; K.4.4
%E.3 Data Encryption
%D.4.6 Security and Protection
%K.4.1 Public Policy
%K.4.4 Electronic Commerce

\newpage

%\part{Bitcoin Technology}

\vskip-6pt
\vskip-6pt
\section{Background: The Emergence Of Bitcoin}
\vskip-6pt

Bitcoin is a collaborative %peer-to-peer
virtual currency and a decentralized
peer-to-peer payment system without trusted central authorities.
%We refer to \cite{NewYorkerBitcoinVeryDetailed} for an introduction to bitcoin.
It has been invented in 2008 \cite{SatoshiPaper} and launched in 2009.
It is based entirely on methods
and ideas already known for more than a decade \cite{DworkNaorJunkMail,WeiDaiBMoney,HashCash}.
The audacity of \cite{SatoshiPaper} was to actually
put these ideas into practice
and design a system able to function for many decades to come.
However,
ever since Bitcoin was launched in 2009,
it has been presented by its technical architects
as an experimental rather than mature
electronic currency ecosystem.
%(Standards vary, but there seems to be a consensus forming around Bitcoin, capitalized, for the system, the software, and the network it runs on, and bitcoin, lowercase, for the currency itself.)
%

In cryptography bitcoin is a practical payment
and virtual currency system.
However from the point of view of financial markets,
it could be considered as a play currency,
an experiment in the area of electronic payment. % which
%Bitcoin implements a particular type of peer-to-peer financial anarchy. % movement.
%Initially a self-governing crypto co-operative and social experiment,
%bitcoins are increasingly traded for real money. %and other commodities,
%%many new business experiments are flourishing
%It became a particularly highly volatile market
%on which 1 billion dollars can %be created or
%evaporate in a matter of hours. % as shown by recent events
%%\cite{NewYorkerBitcoinVeryDetailed}.
%%
This is how it has started:
a %libertarian
self-governing open-source crypto co-operative
and a social experiment.
Initially it concerned only some enthusiasts of cryptography
and computer programmers who believed in it and supported it.
It is not clear whether bitcoin was actually ever meant
to become an equivalent or replacement
of our traditional currencies governed by central banks.
However bitcoin is expected to be a currency or money.
Since 2010 a Japanese company Mt. Gox has started exchanging bitcoins against
real currency in a professional way.
To this day this single company accounts for the majority of such transactions worldwide.
However the press and media coverage and the appreciation of bitcoin on the markets
made that bitcoin is no longer a play currency and is taken much more seriously.

Bitcoin is frequently associated with a criticism of the
sorry state of the global financial industry.
%and its regulation. %Krugman disagrees, all is OK \cite{KrugmanAntiSocialNetwork}.
All bitcoin transactions are descendants of one single initial transaction
which contains a reference to a paper about a government bailout for banks
which have appeared in the British newspaper The Times on 3 January 2009.
In contrast bitcoin does not allow any government intervention
and it is claimed to be immune against inflation. %which such events are associated with.
%It has a built-in very strict control of the monetary supply and
%It is sometimes presented as a sort of silver bullet and a remedy
In this respect bitcoin is sometimes compared to gold \cite{TheEconomistDigitalGold}.

%militant

\newpage
\vskip-6pt
\vskip-6pt
\subsection{Bitcoin Hits The Sky}
\vskip-6pt

\begin{figure}[!ht]%in the log he said he changed !h to !ht, why???
\centering
\begin{center}
\hskip-1pt
\hskip-1pt
\includegraphics*[width=4.9in,height=2.47in,bb=0pt 0pt 782pt 454pt]{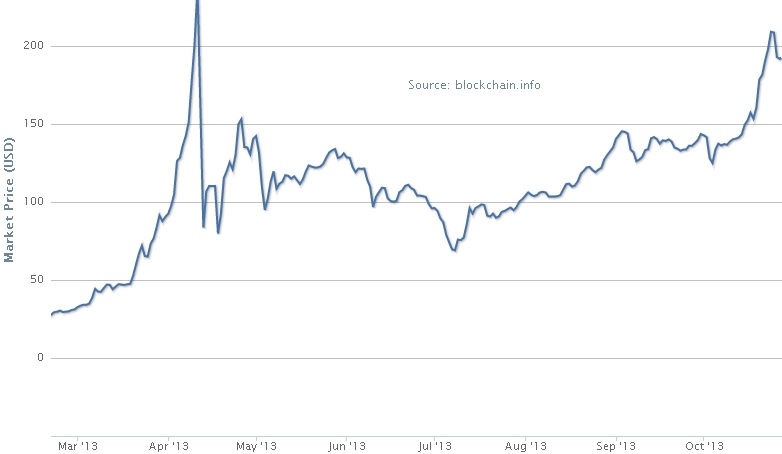}
\end{center}
\vskip-2pt
\caption{
The market price of bitcoin in the last 9 months.
%however the overall main overall effect
%is that
%is simply that they have made it known
%to a much larger number of people.
%even if 80% of people think it is rubbish, 20 % of the much bigger number will buy it and make the share price rise!!!
}
\label{BitcoinExuberanceGraph}
\end{figure}
%\vskip-2pt

In March/April 2013 bitcoin  have attracted a lot of the mainstream press and media
in a very close relation to the aftermath of the bank deposit crisis in Cyprus
\cite{NewYorkerBitcoinVeryDetailed}.
Verbal speculation was accompanied by increased trading and speculation
and the apparent entry of professionals such as hedge funds into bitcoin trading.
%The fact that large investment funds such as hedge funds are moving into this market is maybe hard to hide, see Shamir work..
%%from one blog: ...FT have apparently written that one hedge fund had 20 phone calls per day from clients willing to invest 100 M......
In a space of a few weeks, the market price paid for one bitcoin has more than tripled,
attained about 200 dollars,
then it fell more than 50 percent in a few hours \cite{NewYorkerBitcoinVeryDetailed}.

This event however cannot be seen as a purely financial event:
a major upgrade of Mt Gox exchange web site took place also at this moment.
%Bitcoin have seen its second major and spectacular crash
%and seems to be recovering from it now.
%
%in fact the crisis in cyprus is the wrong reason to interest investors in bitcoin
%many naive investors will not see that the supply of bitcoins have grown so dramatically
%or will think it should be correlated to the increase in market prices,
%while it should be inversely correlated...
%
%\begin{figure}[!h]
%\centering
%\begin{center}
%\hskip-1pt
%\hskip-1pt
%\includegraphics*[width=4.17in,height=2.36in,bb=0pt 0pt 633pt 477pt]{bitcoin130413digitalgold.jpg}
%\end{center}
%\vskip-2pt
%\caption{
%The market price of bitcoin after the crash of early April 2013.
%}
%\label{BitcoinAfter13April}
%\end{figure}
%%\vskip-2pt
%
Likewise bitcoin was very strongly influenced by its growing popularity and media coverage.
Interestingly bitcoin did not fall anywhere near its levels from before March 2013.
On 13 April 2013 the leading newspaper The Economist explains that bitcoin is here to stay,
that it is a future of payments and calls it {\em digital gold} \cite{TheEconomistDigitalGold}.
It is possible to see that the correction which followed
an earlier crash (cf. Fig. \ref{BitcoinExuberanceGraph})
has finished at this precise moment. %and the price could start rising again.
Even though the press and media have actually debated and criticized bitcoin a lot.
%the overall main overall effect is that
%is simply that they have made it known
%to a much larger number of people.
%even if 80% of people think it is rubbish, 20 % of the much bigger number will buy it and make the share price rise!!!
It is like bitcoin has achieved maturity
and become a mainstream financial asset on that very day.
%in this very month of April 2013.
%After the crash however the price remained quite high for many months.
At the moment of writing on 22 October 2013 it has again been tending towards 200 dollars.

\newpage

\vskip-6pt
\vskip-6pt
\section{Bitcoin in Question}
\vskip-6pt

Recent popularity makes that many people ask themselves a lot of questions about bitcoin
which challenges our traditional ideas about money and payment.
%And they get mixed answers.

\vskip-6pt
\vskip-6pt
\subsection{Is Bitcoin a Currency?}
\vskip-6pt

Some commentators explain that bitcoins function essentially like any other currency
\cite{NewYorkerBitcoinVeryDetailed}.
The Economist calls it a currency, digital money, and compares it to {\em digital gold}
\cite{TheEconomistDigitalGold}.
Other authors, such as Paul Krugman, Nobel price in economics,
have many times criticized bitcoin
as just one of possible ways to pay electronically \cite{SummaryKrugman}
which would not have all the desirable characteristics of a modern currency
\cite{SummaryKrugman,KrugmanAntiSocialNetwork}.
According to the Forbes magazine bitcoin
is not money because it does not have a stable intrinsic value \cite{SteveForbesBitcoinsNotMoney}.
%and has rather risky and uncertain prospects.
%which question we also consider
%% full version %%In the present paper we will also arrive at similar??? conclusions.
However all these arguments seem to be rather superficial.
There are much more serious questions at stake.
%It seems that nobody have yet seen what the real argument is.
Money has benefited from extremely {\bf strong legal protection} in most countries for centuries.
This is combined with effective {\bf policing of fraud} which costs a lot of money.
Major governments with their armies stand behind their money.
%%%Criminals sometimes counterfeit currency, foreign governments could do it without any problem. International law prevents???
All this makes that counterfeit currency rates with coins and paper money %%%not at all and in the traditional retail banking
are surprisingly low.
They are roughly just 1000 times smaller
compared to higher rates of fraud which are systematically seen
with modern payment technologies:
for example we have known massive amounts of bank card fraud
such as skimming.
%EU position?
%There is an inherent cost of running any payment system and any monetary system, Krugman is wrong
%Krugman is wrong about the gold standard and environmental crictics are wrong...
%Money is not just a service or a technology or a product,
%it a social and political fact which
%underpins and supports the economy.

In this paper we will present some additional arguments to the effect that
bitcoin cannot (yet) in the current state be called a currency,
mainly due to its inherent built-in instability
cf. Section \ref{KillerBusinessCycleArgument}.
%% full version %%and Fig. \ref{BitcoinMarginSuperMisleadingChartROS}.
These properties however, could be eventually fixed.
% and it would indeed work like a currency.
%though in practice it is likely to be problematic,
%see Section \ref{KillerBusinessCycleArgument}

\vskip-6pt
\vskip-6pt
\subsection{Some Interesting Good Points About Bitcoin}
\vskip-6pt

Not all is bad with bitcoin.
Bitcoin has a lot more to offer than
prospects to make or lose a lot of money
in yet another high-tech investment scheme,
with interrogations whether it is just another Ponzi scheme.
It has some very interesting properties:
it allows payments to be sent in a short time.
Payments are irreversible.
It offers some degree of %unprecedented and
%genuine
public verifiability, and transparency
for %the source code - will be contradicted ... and
all bitcoin transactions without any exception.
Bitcoin transactions can to some extent be traced
and connected to each other, see \cite{ShamirBitcoinGraph}.
It could be regulated by governments like any other financial market.

Bitcoin shows that with modern technology it may radically
cheaper than previously thought to run an electronic payment system.
The cooperative and democratic character of the bitcoin currency
should be a model for the financial industry
which has lost a lot in confidence of the public.
%It offers pseudonymity rather than full anonymity %which is almost impossible to achieve
% in the US , the Financial Crimes Enforcement Network (FinCEN) announced guidelines requiring certain "virtual currency" trading entities to register as Money Services Businesses (M.S.B.s).
%"virtual currency" trading entities to register as Money Services Businesses (M.S.B.s).

More importantly bitcoin
%implements
has a major built-in feature: %protective mechanism against inflation
%feature: %innovation
%in the history of money and monetary policy:
an artificially limited and strictly controlled monetary supply.
This built-in deflationary scarcity
%explains why bitcoins appreciate in value and
contrasts with
the %worryingly %inflationary %excessive %creeping
potentially unlimited monetary %supply
expansion
%inflation
%erosion of the %purchasing power ? face value ?
%in the quantity of
of major traditional currencies %in circulation
which have been observed in the recent years.
This has been praised by many commentators
and at occasions bitcoin is presented as a miracle remedy against inflation.
The Economist have famously compared bitcoin to {\em digital gold}
\cite{TheEconomistDigitalGold}.

Unhappily as we will see in this paper,
if we look at the details,
we will see that this mechanism of decreasing returns
is designed in a highly problematic way,
cf. Section \ref{KillerBusinessCycleArgument}.
This for no apparent reason.
We will also explain how to fix this problem.

\vskip-6pt
\vskip-6pt
\subsection{Bitcoin Fundamentals}
\vskip-6pt

It is not true that bitcoins have no intrinsic value \cite{SteveForbesBitcoinsNotMoney}.
On the contrary.
A lot value comes from the network effects \cite{OdlyzkoMetcalfe}.
The network of bitcoin nodes has acquired some serious value.
The network of bitcoin supporters and users and their faith in bitcoin
adds more value.
Its popularity is astounding.
The easiness with which payments can be made within this system is valuable.
The security properties of how hard it is to make bitcoins,
steal them, or create new bitcoins, have some price. %are of great value.
All these properties add value to bitcoins.
In general the value of bitcoins comes from distinct sources than that of the
traditional financial instruments, but one cannot possibly deny that bitcoin
has some value, which is considerable if we look at it the current market price.

Moreover this value grows with time as the network grows.
It is widely known that the value of networks grows faster than linearly.
Networks are expected to generate value which is more than the sum of the parts
which constitute the network \cite{OdlyzkoMetcalfe}.
The particularity of the bitcoin network
is that it has accumulated an astronomical amount of
``cryptographic evidence'' which testifies about the whole
bitcoin history in detail,
and which it would be very hard and extremely costly to
%reproduce, imitate or
falsify.
% in the form of distributed computationally-intensive tasks
%which has created chains of nested transactions,
%It would cost at least one hundred of millions of dollars to re-do this work.
In this aspect bitcoin performs all the desired functions
of an electronic notary system which is there to certify its own past history.
Such a system is also valuable.

In this paper we look at bitcoin from the point of view
of a curious cryptologist and an information security expert.
Ideally we would like to see if bitcoin is secure
and what kind of attacks are possible against bitcoin,
and maybe how it can be improved, cf. \cite{BitcoinFC12SecurityOverview}.
However we cannot pretend to cover all these questions in one paper.
Instead we concentrate just on the questions of bitcoin mining.
%% full version maybe %%and whether what is generally believed
%% full version maybe %%about bitcoin and bitcoin mining is true.
%% full version maybe %%This from a pragmatic and cautious security perspective.
%which teaches us never to underestimate the attacker.
We look at the cryptographic foundations of bitcoin
and also consider some potential broader
business and financial consequences
of these questions.
In particular we look at various factors which are likely to affect the stability
of bitcoin as a currency in very substantial ways and at additional factors
which will affect the bitcoin market participants and investors.
%%%%%%%%%%%such as questions of trust.

\newpage

\part{Bitcoin and Bitcoin Mining}
\label{Part1AllAboutMinining}

\section{The Challenge}

Each time a certain type of good, medium or technology
is widely adopted %(in human history or more recently)
as a mean of exchange and payment,
and regardless if we would really agree call it ``a currency'',
it has certain pre-existing characteristics.
These are its (relative) rarity,
security understood as important difficulty to forge this ``currency''
and/or to commit fraud.
%and in the case of government-issued currencies, legal protection and
%highly effective policing of fraud and crime.
The rules which govern the adoption of cryptographic technology are not dissimilar.
%another type of security technology.
In cryptography, when a certain cryptographic scheme is
massively adopted, it is so because it is hard to break.
%all code breakers on this planet have tried to break it and failed.
%Such cryptographic schemes are also relatively rare and precious.
Money has somewhat to resist fraud theft and forgery,
cryptography has to withstand the presence of hackers and code breakers.
Both emerge through the process of {\bf natural selection} in which
different types of payment technology or/and different
sorts of cryptography co-exist.
With time some solutions emerge as a preferred choice
however a certain bio-diversity always remains.
%compete against each other

In this paper we study bitcoin with particular attention paid to the process of bicoin mining,
which is the specialist term given to the process by which new bitcoins are created.
We are going study this question from many different angles.
It is a cryptographic puzzle,
but also a disruptive technology in monetary history.
It also is method to make money for miners, a method to own and control the bitcoin currency,
a method to police the bitcoin network and enforce the compliance with a certain version of the bitcoin specification,
etc.
%% full version %%We also study some business questions such as one of mining profitability
%% full version %%and comment on recent history of mining.
%certain quite specific sorts of moral dilemma.
Later in Part \ref{PartImprovedMinining}
we are going to
to study in a lot of detail
one particular technical question
in %pertaining to
symmetric cryptography
to see if there exists an improved method
which allows one to mine bitcoins faster.

\section{What Are Bitcoins and Bitcoin Mining}

Bitcoins are a type of digital currency which can be stored on a computer,
though it is advisable to store them rather in a more secure way.
For example on paper and in a safe,
or on a smart card or another highly secure platform.

Bitcoins use the concept of
so called ``Proofs or Work''
which are solutions to certain very hard
cryptographic puzzles based on hash functions.
However these solutions are NOT bitcoins.
The puzzles are rather part of the bitcoin trust infrastructure.
In fact the puzzles are connected together to form a chain
and as the length of this chain grows,
so does the security level.
%called Reusable Proofs of Work (RPOW)
%and
%digital signatures which attest their
Bitcoins are simply awarded to people who produce these ``Proofs or Work''
which is a very difficult task.

Ownership of bitcoins is achieved through digital signatures:
the owner of a certain private key
is the owner of a certain quantity of bitcoins.
This private key is the unique way to %exercice ownership, for example to
transfer the bitcoin to
another computer or person.

The operation of so called {\em bitcoin mining}
or creating bitcoins out of the thin air
is not only possible.
It is essential, it is encouraged, and it is a crucial %
and necessary part of the Bitcoin ecosystem.
%mining could be illegal? see \cite{NewYorkerBitcoinVeryDetailed}
%Since mining yields pocket change for most, even if it were technically a violation of the way FinCEN sees the law, mining without registering would be like "laundering" a twenty-dollar bill by taking it to the grocery store and asking for two tens… it's hardly worth the resources for anyone to care about it, no matter how illegal they decide it should be.
Cryptographic computations executed by a peer-to-peer network of a growing
network of currently some twenty thousand independent nodes \cite{NewYorkerBitcoinVeryDetailed}
are the heart of the security assurance provided by this virtual currency system.
It would be very difficult and extremely costly
for one entity to corrupt all these independent people.
%One billion dollars could be not enough to corrupt this bitcoin ecosystem.
The sum of all this collective computational work provides
some sort of solid cryptographic proof
and prevents attacks on this system.
This also how the network polices itself:
miners are expected to approve only correctly formed transactions.
Bitcoin implements a specific sort of distributed
and decentralized electronic notary system without a central authority.
Well almost.
Certain decisions about how the system works,
what exactly the bitcoin software does
and how \cite{BitcoinMainSoftwareDistribution},
are still pretty centralized.
They are subject to adoption or rejection by the wider community.

% now 25 bitcoin for each solution, to ba halved many times...
%In the future, as the number of new bitcoins miners are allowed to create in each block dwindles,
%the fees will make up a much more important percentage of mining income.

In a nutshell,
bitcoin miners make money when they find a 32-bit value which,
when hashed together with the data from other transactions with a standard hash function
gives a hash with a certain number of 60 or more zeros.
This is an extremely rare event.
It is in general believed that
there is no way to produce these data otherwise than by engaging in very long
and costly computations.
This question of feasibility of bitcoin mining and possible improvements
is a central question in this paper
and we study it later in more details.

%\newpage
\section{Are Bitcoins Secure?}

Is Bitcoin a secure distributed system
and in what sense it is secure remains unclear.
As far as we can see nobody yet claimed that bitcoin is provably secure as we understand it in cryptography
with a formal definition and a security proof. %of what it means for this system to be secure.
On the contrary, {\bf bitcoin clearly isn't a state of the art cryptographic system},
see \cite{BitcoinFC12SecurityOverview}.
It is a practical system with many potential shortcomings.
In this respect it has been a tremendous success and
it has no serious competitor at the present moment.

For the time being we need to assume that
the security of bitcoin payments is based on the shared
belief that there is no way to hack the bitcoin system in any substantial way.
Officially bitcoin is experimental, it does not claim to be secure.
In fact the security of the bitcoin protocol and software
{\bf has already been broken once}.
On 15 August 2010
somebody has created an unbelievably high quantity of
184 billion of bitcoins worth literally trillions of dollars
out of nothing and made the distributed system accept it,
see \cite{BitcoinExposuresList}.
The protocol system and software
have been patched immediately
and bitcoin protocol is now at version 2.
All bitcoin adopters worldwide had to agree to discard
this strange attribution of money.
The only way to recover from this sort of error is by consensus.
There were also major cyber attacks
with concrete exploits against bitcoin software and systems,
see \cite{BitcoinExposuresList}.
In just once such incident 17000 BTC were lost
(or maybe stolen) which is worth millions of dollars.
%by a Polish exchange under attack
%01 August 2011
%http://siliconangle.com/blog/2011/08/01/third-largest-bitcoin-exchange-bitomat-lost-their-wallet-over-17000-bitcoins-missing/
These embarrassing incidents are not very widely publicized.

Moreover there are some non-technical reasons to be very cautious with bitcoin.
Quite interestingly the creator of the system \cite{SatoshiPaper}
was apparently a pseudonym and seems to have disappeared.
As far as we can see no serious academic cryptologist
has publicly expressed their faith in bitcoins and their security.
%There is not enough people who scrutinize the security of bitcoin
%\cite{BitcoinExposuresList,BitcoinFC12SecurityOverview}.
On the contrary, the cryptographic community,
as well as the software engineering community,
are full of highly capable code breakers
able to find new attacks and exploits on secure systems
such as bitcoin every day.
%A good record track is not a a substitute for security.
%past history of limited ...

A major reference in this area is a paper published at Financial Cryptography 2012
\cite{BitcoinFC12SecurityOverview}.
This paper clearly explains that
{\em hundreds of academic papers have been published to improve
the efficiency and security of e-cash constructions}.
At the same time the authors explain that bitcoin
is a rather simple system which
{\em uses no fancy cryptography, and is by no means perfect}.
Then they analyse the security of the bitcoin system
from numerous angles
and consider many interesting attacks,
see \cite{BitcoinFC12SecurityOverview}.

In this paper we
look mostly at the questions of
how bitcoin works and how exactly bitcoin mining works.
We try to see if it is possible to improve this process to be more efficient.  %
%and the benefit vs. cost analysis. % of bitcoin mining.
Later we will look at what are the consequences of what we have learned.
%% full version maybe %%We try to go beyond common ideas about bitcoin and try to challenge these ideas.

%\newpage

\vskip-5pt
\vskip-5pt
\section{The Main Activity of Miners}
%Key Hard Problem:
%How Difficult Is It To Mine Bitcoins?}
\label{MainGoalsAndActivityOfMiners}
\label{StaticCISOProblem}
\vskip-5pt

The goal of the miner is to solve a certain cryptographic puzzle
which we will later call {\bf a CISO Hash Problem}.
The solution will be called {\bf a CISO block}.
Great majority of miners
ignore what exactly  they are doing, they are running either
open source software or have purchased some hardware to do mining very efficiently.
However miners must know that the operation is very timely and that they need
to be permanently connected to the network.
The solutions to these puzzles are linked to each other
and form a {\bf unique} chain of solutions.
This is usually called {\em the block chain}.
The whole block chain is published on the Internet.
The whole of it can for example be consulted at \url{http://blockexplorer.com/}.
All new blocks which are found need to be broadcast to all network participants as soon as possible.
The miners need to be very reactive and they do it because it is in their interest
(note: a very recent paper proposes
another strategy \cite{BitcoinBrokenGrowingSelfishPoolStrategy}).
They need to listen to broadcasts in order to receive the data about recent transactions which they are expected to approve.
Then they need need to broadcast any solution (a CISO block) which they have found as soon as they found it,
because their solution is likely to be part of the {\em "main chain of blocks"} only if it is widely known.
Once the solution is known it "discourages" other miners from searching for the same block.
Instead they can concentrate on searching for the next block which will confirm the present block
and will make the miner be able to claim hist a reward for producing this  CISO block.

Our goal is to clarify how this system works.
In the present section we consider a static
computational problem which needs to be solved.
In Section \ref{MovingTargetCISOFurtherExplanation}
we will further explain the dynamics of bitcoin
production in the long run:
how this problem changes with time
in a predetermined way. % and what are the consequences of this.

\subsection{Bitcoin Mining vs. Block Cipher Cryptanalysis}

The problem of bitcoin mining is very closely related to well known problems in cryptography.
%Moreover very substantial budgets of tens of millions of euros have been spent
%on the study of these questions in the past.
%It would be very surprising that
%this would not affect the bitcoin market in some way.
%
One crucial question is as follows:
how does the bitcoin mining differ from traditional questions in
cryptanalysis of block ciphers and hash functions
and is there a more efficient way to mine bitcoins.
%This question requires further research.

First we are going to briefly describe the problem as a static computation problem
about a certain block cipher.
Then we are going to look at how the problem evolves in time
and how solutions to the CISO problem are converted to shares in the bitcoin currency.
Finally we are going to study what the possible solutions and optimizations are.

\vskip-5pt
\vskip-5pt
%\subsection{The Description of the Problem}
%\subsection{The Reduced Space Hashing Problem}
%\subsection{Constrained Output Hashing (COH) Problem}
\subsection{Constrained Input Small Output (CISO) Hashing Problem}
\label{FirstQuickDefCISOProblem}
\vskip-5pt

New bitcoins can be created if the miner can hash some data from the bitcoin
network together with a 32-bit random nonce, and obtain
a number on 256 bits which starts with a certain number of zeros.
We call this problem CISO:
{\bf C}onstrained {\bf I}nput {\bf S}mall {\bf O}utput.

This can be seen as a special case of
a problem which is sometimes called CICO
which means
{\bf C}onstrained {\bf I}nputs {\bf C}onstrained {\bf O}utputs problem.
This terminology have been introduced recently
in the study of the most recently standardized U.S. government
standard hash function SHA-3 a.k.a. Keccak.
SHA-3 is the latest hash function in the SHA family
and a successor to SHA-256 used in bitcoin \cite{FirstAnalysisKeccak}.
It is possible to claim that
this means that SHA-256 of Bitcoin is
considered by the United States NIST
and a broader cryptographic engineering community as
NOT sufficiently secure for long term security.
This sort of CISO/CICO problems are not new,
they are very frequently studied in cryptanalysis
of hash functions since ever, and endless variants of these problems
exist for specific hash functions, some examples can be found in
\cite{FirstAnalysisKeccak,gosthashCourtois,MatusiewiczModifiedSHA256}.

The exact details of the specific
Constrained Input Small Output (CISO) problem
which we have in bitcoin are described below.
It can be obtained by the inspection
of the bitcoin source code,
see \cite{BitcoinMainSoftwareDistribution,OpenSourceMiningApps}.
Both code for bitcode mining and for
the whole bitcoin network is open
and therefore the process
is relatively transparent.

%The difficulty is that in cryptography research
%almost always sooner or later we will discover
%that there might a better
%and more profitable way to do the same thing.

\subsection{(CISO) Hashing Problem Internals}
\label{CISOProblemInternals}
\vskip-5pt

\begin{figure}[!h]
\centering
%\vskip-2pt
\begin{center}
\hskip-1pt
\hskip-1pt
\includegraphics*[width=5.1in,height=4.2in,bb=0pt 0pt 1180pt 840pt]{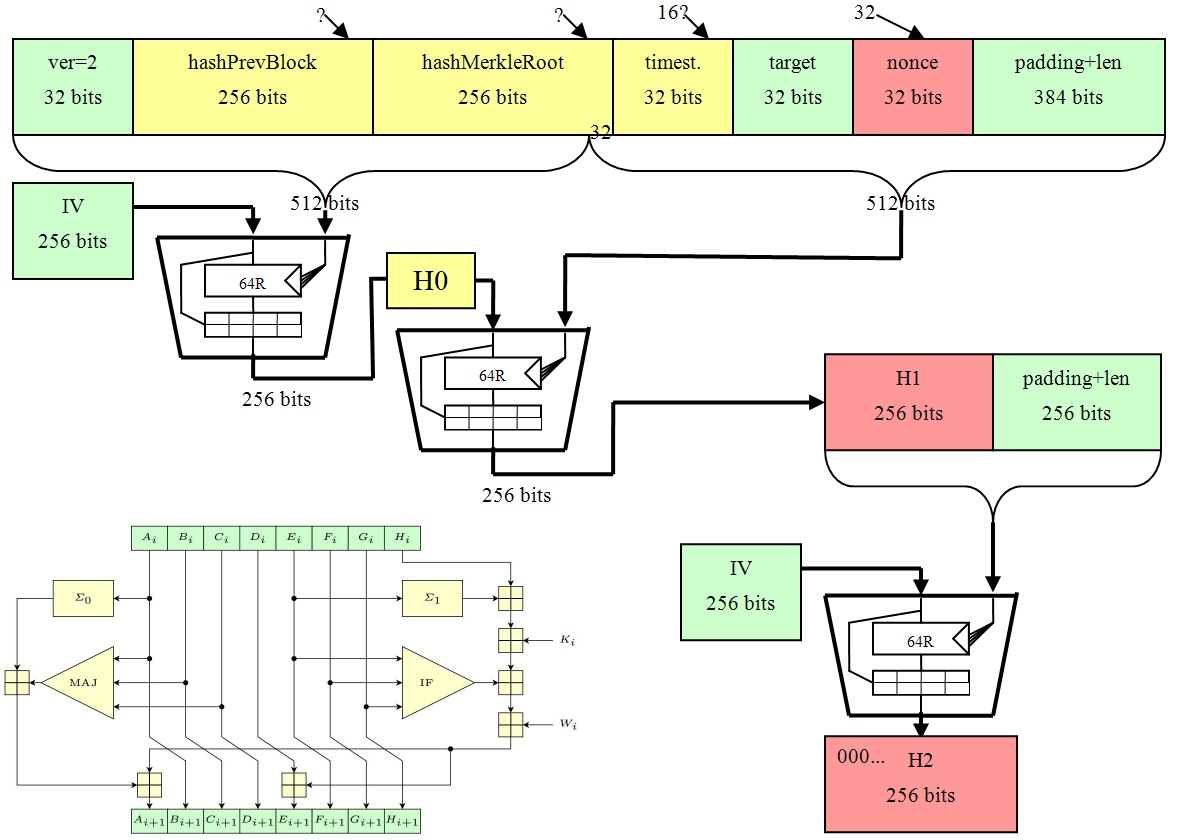}
\end{center}
\vskip-2pt
%\caption{Internal structure of the first SHA-256 application used in bitcoin mining}
\caption{
The Block Hashing Algorithm of bitcoin revisited and seen as a
Constrained Input Small Output (CISO) problem.
We see two applications of SHA-256 together with
internal details of the Davies-Meyer construction. %, as used in bitcoin mining.
We can view it as a triple application of a specific block cipher.
%The standard process of bitcoin mining is very similar to a brute force attack on a block
%cipher with a specific very particular objective to attain.
An interesting question is whether there is a more efficient
cryptographic shortcut or inversion attack or some non-trivial optimizations
which allow to save a constant factor.
Such optimizations, if they exist,
could be worth
%millions of dollars
some serious money
as they would allow to produce
bitcoins cheaper.
%which would achieve the same objective at a lower cost.
%see Fact \ref{HashSpeed1.8X}.
% with 256 bit block size,
%64 rounds, and 512 bit key size expanded to 64 subkeys on 32 bits each.
}
\label{BitcoinCISOProblem}
\end{figure}

On Fig. \ref{BitcoinCISOProblem}
we show the cryptographic computation
which is %repeatedly
executed many many times by bitcoin miners.
This picture emphasizes the internal structure
%hidden
inside SHA-256 hash function.
The inputs and contraints on these inputs are explained
in details in Section \ref{InputsExplainedSection} below.

%Mining consists in repeatedly computing hashes of variants of a data structure called a {\em block header},
%until one is found whose numerical value is low enough.
%\cite{RosenfeldPoolRewardPaper}

%\newpage

SHA-256 is a hash function built from a block cipher following
the so called Davies-Meyer construction.
The principle of the Davies-Meyer construction is that the input
value is at the end added to the output and that it transforms an encryption algorithm
into a ``hashing'' algorithm, a building piece of a standard hash function.
The underlying block cipher has 64 rounds and thus a 2048-bit expanded internal key
(64x32 bits).
This key is obtained from the message block to be compressed,
which has 512 bits at the input and is expanded four times
to form this 2048-bit internal key for our block cipher.
In one sense on Fig. \ref{BitcoinCISOProblem} we
convert the problem of bitcoin mining
or of solving CISO hash puzzles,
to a specific problem
with three distinct applications
of the block cipher
which underlies SHA-256 connected together to form certain circuit.
%We refer to leading experts of SHA-256 analysis
%in the cryptographic literature,
%see for example \cite{SHA256PreimageMatusiewicz,MatusiewiczModifiedSHA256}.
%The block size in this block cipher is 256 bits,
%the key size is 512 bits which is expanded to 64 subkeys on 32 bits each
%for each of 64 rounds of the cipher.
More details about these inner workings
of SHA-256 as it is used in Bitcoin mining
will be given in Section \ref{DescriptionOfHashing1}.

%\vskip-2pt

\vfill
\pagebreak

\vskip-5pt
\vskip-5pt
\subsection{The Main Objective of CISO Hashing}
\label{TargetExplained}
\vskip-5pt

The goal of CISO hashing is to produce solutions which
are correctly formed in the sense that they satisfy
all the required conditions and constraints,
which we are going to explain in details in Section \ref{InputsExplainedSection}.
The miner is trying to find a solution to the CISO problem such that

$$H2 < \mbox{\var{target}}.$$

Here \var{target} is a large integer which is a global variable
for the whole bitcoin system worldwide,
and on which all
the participants worldwide are expected to agree.
The value of  \var{target} slowly changes with time
and is adjusted approximately every 2 weeks.
More details are given below in Section \ref{MovingTarget}.

The job of bitcoin miners is to find these solutions and publish them.
They are rewarded with some bitcoins for their work.
In 2013 the reward is 25 BTC (25 bitcoins) per valid solution.
How exactly this reward works and
how it changes over time will be explained later.

It is generally believed that there is no other method to achieve success
than trial and error; hashing at random
as depicted on Fig. \ref{BitcoinCISOProblem} until
a result with a sufficient number of leading zeros is found.
%contrived instances could be be easily detected!!! counter-cryptanalysis!!
%The main objective of this paper is to see if this belief is reasonable.
However this is unlikely to be true, there is always a better way,
at least slightly,
see Section \ref{SpeedFinalResult1.8X}.
%and this topic requires further research.

\vskip-5pt
\vskip-5pt
\subsection{The Inputs and Input Constraints In CISO Hashing}
\label{InputsExplainedSection}
\vskip-5pt

A number of data fields are present as inputs to the CISO problem,
cf. Fig. \ref{BitcoinCISOProblem}.
Some data are fixed or change very slowly and these are indicated in green on Fig. \ref{BitcoinCISOProblem}.
Other data need to be adjusted by the miner in order to obtain a solution however they still change very slowly.
Such data are indicated in yellow on the main picture.
Data which change the most frequently are indicated in red: these are
``hot'' data which need to be re-computed each time the nonce changes.

Bitcoin is a live distributed system and
the exact conditions required for the data to be consider valid
are essentially fixed and known, however
they are likely to evolve with time in subtle ways.
Rules have already been and are likely to be altered
during the operation of the system.
They also depend on the consensus of participants in the system.
It is generally admitted in the bitcoin community that
there could be  and should be many different versions of the software which co-exist.
This is because if all the software came from one single source,
bitcoin would cease to be a system independent from any central authority and
would develop a serious syndrome of a single point of failure.
Therefore bitcoin software should be diverse.
We could even postulate that nobody should be excluded
from making their own software even though we might be worried about
Denial of Service attacks.
In practice the reality is different:
the original Satoshi software \cite{BitcoinMainSoftwareDistribution}
conserves a prominent position.

In what follows we are going to describe
what different inputs are.
We need to pay attention to the degree of freedom which is allowed:
to what extend the miner is able to select different values
in order to achieve the desired result.
We have:
%Here is what the inputs are:

\vskip-5pt
\vskip-5pt
\begin{enumerate}
\item {\bf Version number on 32 bits. }
This is an integer which represents the version number of the bitcoin software.
It defines the rules which govern the blocks, which blocks could be accepted as valid.
It is essentially a constant,
since the creation of the system in 2009 it has always been equal to 1 then it became 2.
At the time of writing
%nearly $90 \%$ of
new blocks are typically version 2 and
%CHECK this???????????????????????????????????????????????????
it has been announced that very soon the community
will stop accepting blocks generated compliant with older version 1 rules.
%the moment and the upgrade .
%https://en.bitcoin.it/wiki/BIP_0034
%Motivation
%Clarify and exercise the mechanism whereby the bitcoin network collectively consents to upgrade transaction or block binary structures, rules and behaviors.
%Enforce block and transaction uniqueness, and assist unconnected block validation.

\item {\bf The previous block on 256 bits. }
Or more precisely a hash on 256 bits of all the data of the previous CISO block encoded in a specific way.
Each new block is added at the end of the chain of blocks.
Ideally there is only one official chain of blocks.
%In practice there may be

The miner has essentially no choice,
a new CISO block is created approximately every 10 minutes
and it is broadcast in the peer to peer network (and published on the Internet)
as soon as possible.
The current block very quickly becomes obsolete.
Either miners will now use it as a previous block,
or they will use another freshly generated solution.
The solution is NOT unique however there is only one winner (in the long run).
Each miner who produces a solution wants this solution to be known by the highest possible
number of other miners and ahead of any competing solution.
%The miner needs to solve the current CISO block puzzle of Fig. \ref{BitcoinCISOProblem}
%precisely for the last available previous CISO block and he wants to be first
%to achieve this result.
It is a race where every microsecond counts.

The process of generating new blocks is a sort of lottery
and the probability that there will be two winners in a very short interval of time is low.
It is the miner's responsibility to check very frequently
if a solution have not already been found.
If this is the case, it is not in his interest to find another one later on, his chances to succeed decrease quickly with time.
On the human and technical/software side however,
network propagation makes that not all participants have the same view of which solution was the first,
and there is a real possibility to have disputes.
Potentially bitcoin could split into two systems
which do not recognize each other and which operate two independent ever growing blockchains.
%
%In theory bitcoin could function in a fully decentralized way where different solutions
%could co-exist for a very long time and there could be a split in the system
%with two block chains which do not recognize each other and do not want to agree. %in terms of priority.
%In practice a number of trusted people and companies hold the community together.
%%really? maybe discard if same timestamp? Maybe 2nd solution is easier?
In theory there is a so called Longest Chain rule
which allows to solve this problem \cite{SatoshiPaper}.
In practice it is a bit different. The Longest Chain rule might be hard to enforce.
People may rather trust a well-established website than siome votes which come from the
(more obscure) peer-to-peer network.
They could suspect or resist an attack by a powerful entity.
%(which thing could allow effectively to cancel past transactions and double spend).
They can simply agree to disagree because they have spent
some substantial money on electricity
on one version of the chain.
%% full version %%We discuss this question more in details in Section
%% full version %%\ref{LongestChainMajorityIsItTrue} below.

%In practice the time is variable,
%see Section \ref{MovingTarget}.
%The system is regulated in such a way that it always is about 10 minutes
%simply by adjusting \var{target} up and down which is updated once
%approximately every two weeks.

\item {\bf The Merkle root on 256 bits}.
This is a sort of an aggregated hash of many recent events in the bitcoin network
which certifies that the system recognizes all of them simultaneously as being valid.
Moreover it also has a {\bf self-certifying property}.
It also hashes and certifies the public key
of the future owner of this freshly created portion of bitcoin currency
which this block is intended to embody,
as soon as it is in turn confirmed by few other CISO blocks.

The current CISO block which the miner is trying to create by solving the current
CISO problem and all subsequent CISO blocks will provide an accumulation of evidence
about all these events which will be increasingly secure and increasingly hard to falsify with time.
This security guarantee increases %accrued
with time.
It is achieved because miners expend a lot of computing power,
the number of miners is increasing,
and more recently they use specialized devices with increasing fixed (equipment) costs.
All these things are increasingly hard to imitate.
%amazing, it is like postponing the proof, see Zero-Knowledge Fisher-Micali-Rackoff

Interestingly the value of the Merkle root can (and needs to be) influenced by the miner as explained below.
First of all it is clear that miners do have some discretionary powers %here
and should be able to collude and/or select
which transactions are going to be recognized by the system.
However the transaction fees which are decided at the moment of making a transfer from
one bitcoin address to another are an incentive to include every single transaction.
Thus the miner is able to collect hundreds of transaction fees for each block generated.

The Merkle root value is always produced by a hash function
which is expected to behave as a random oracle.
This means that the miner can influence this Merkle root value but only very slightly,
basically by trial and error.
%Some values might

\item {\bf Timestamp on 32 bits}.
This is the current time in seconds.
The miner can hardly change it.
This would be extremely risky as another solution could be submitted
at any moment. There are only 600 seconds in 10 minutes.

\item {\bf Target on 32 bits}.
More precisely the global variable \var{target} is on 256 bits
and what is stored here is a compressed version of \var{target}
which is frequently called $\mbox{\var{difficulty}}$.
We have $\mbox{\var{difficulty}} \approx
%8,974,296$
%21,335,329\approx 2^{24.4}$ %at the moment of writing on 7 July 2013.
267,731,249\approx 2^{28.0}$
%\url{http://bitcoindifficulty.com/}
as of 22 October 2013.
This \var{difficulty}
is a real (floating) number which is at least 1
and which is stored in a 32-bit format.
%which encoding is sometimes also called $\mbox{\var{header.bits}}$.
We have
$\mbox{\var{difficulty}} \cdot 2^{32}
=  1 / \mbox{\var{probability}}
=  2^{256} / \mbox{\var{target}}$.

%This value \var{target}
%typically  goes up and sometimes slightly down in the system
%in order to insure that CISO blocks are produced at a predetermined speed
%of one every 10 minutes,
%see Section \ref{MovingTarget}.
%The exact value \var{target} changes roughly every two weeks and
%from the point of view of the miner
%at any given moment should be considered as a constant.

\item
{\bf Nonce on 32 bits}.
This nonce is freely chosen by the miner.
Interestingly the nonce has only 32 bits while the
current value of \var{target} makes that
the probability of obtaining a suitable $H2$ by accident
is as low as $2^{-60.00}.$
%difficulty=00000000000000100AB600000000000000000000000000000000000000000000   %22 Oct 2013
%convert("00000000000000100AB600000000000000000000000000000000000000000000", decimal, hex);
%     100696259189502783924473792493100546893980348528488767029248
%=2^{196.00}=256-

This means that the miner needs to be able to generate different versions
of the puzzle with a different Merkle root
(or with other differences)

\item {\bf Padding+ Len has 384 bits for H1 and 256 bits for H2}.
These are two constants due to the specification of SHA-256 hash function
which is used here twice with data of different sizes:
the input hashed has 640 and 256 bits respectively in each application of SHA-256.
These two values never change.
\end{enumerate}
\vskip-5pt

%=> was already explained before long enum...
%It is very hard to define these things precisely
%because we are dealing with a live distributed system and
%{\bf rules have already been
%and are likely to be altered
%as it is in operation}.
%They also depend on the consensus of participants in the system.
%We can only try our best.

With respect to the input data requirements and constraints above
and the output constraint
$H2 < \mbox{\var{target}}$ we have:
%% full version maybe %%the following very closely related definitions.

\vskip-5pt
\vskip-5pt
\begin{defi}[CISO Problem]
We call the CISO Problem
the problem of finding a valid Merkle root
and other data as
illustrated on Fig. \ref{BitcoinCISOProblem}
which %in absence of another solution ????
is correctly formed and will be accepted
by the majority of current bitcoin software.
\end{defi}
\vskip-5pt

%\newpage

\vskip-5pt
\vskip-5pt
\section{Evolution of The Difficulty of CISO Puzzles With Time}
%\section{Evolution of Bitcoin Mining Task With Time}
%\section{Bitcoin Mining In the Long Run}
\label{MovingTargetCISOFurtherExplanation}
\vskip-5pt

The integer \var{target} is system-wide variable
which should be the same for all bitcoin miners
at any given moment. From the point of view of the miner
it should be considered as a constant.
It changes %roughly speaking every two weeks
with time according to pre-determined rules.
It determines how hard it is to solve the CISO problem.
It is implements self-regulation.
The \var{target} value is adjusted simply in such a way
that the total number of CISO blocks found by all miners on our planet
taken together is constant in a given period of time.
%The total number of blocks generated however is unlimited and grows with time.

\vskip-5pt
\vskip-5pt
\subsection{Moving Target - Regulation Of Speed At Which CISO Blocks Are Generated}
\label{MovingTarget}
\label{SpeedRegulation}
\vskip-5pt

More precisely and according to the rules
embedded in current bitcoin software,
on average one block should be generated
every 10 minutes.
This regulation is achieved by observation
of the speed at which shares have been
generated in a fixed period of time
and adjusting the global variable \var{target} accordingly.
The exact value \var{target}
changes roughly every two weeks,
or when exactly 2016 new blocks have been produced.
It goes without saying that the actual speed at which the blocks are generated is not uniform,
however it remains close to uniform.

As of 22 October 2013 we have
$$\mbox{\var{target}} \approx 2^{256-60.00}.$$
%or equivalently H2 seen as an integer must start with 60 zeros in binary.
This current probability of $2^{-60.00}$ corresponds to the requirement of
having 60 leading zeros, and in fact slightly more or less slightly less than 60,
the exact rule is simply that the equation $H2 < \mbox{\var{target}}$ needs to hold.
The value of  $\mbox{\var{target}}$ changes quite frequently.
%
%difficulty=00000000000000100AB600000000000000000000000000000000000000000000   %22 Oct 2013
%convert("00000000000000100AB600000000000000000000000000000000000000000000", decimal, hex);
%     100696259189502783924473792493100546893980348528488767029248
%=2^{196.00}=256-

This current probability of $2^{-60.00}$
is really {\bf the} success probability for mining
by hashing the appropriate data at random with a double application of the hash function
(as required cf. Fig \ref{BitcoinCISOProblem}).
This is already an extremely small figure.
It makes bitcoin mining very difficult.
It reflects the fact that many people have already solved
many CISO puzzles and obtained bitcoins as a reward.
Tens of thousands of people worldwide mine bitcoins
with increasingly powerful computing devices.
Accordingly the difficulty increases
which is necessary in order to keep limited monetary supply in the system.
At the moment of writing the global hash rate in the network was already %nearly
3000 TH/s and has increased about 50 times in the previous 6 months.

{\bf Remark 1.}
Typically \var{target} decreases with time.
%It is frequently assumed that typically \var{target} will always decrease.
%This cannot be true.
However it can also go slightly up
in order to insure that CISO blocks are produced
at a uniform speed.
It should and could substantially increase if
for some reason the global production of CISO
blocks goes down,
for example due to increased electricity prices or
important ``real money'' capital outflows
in the bitcoin market.
Such events are more than likely to happen,
see Section \ref{TheBigCycle}.
% and bitcoin businesses.

%As already explained, at the moment of writing
%we have $\mbox{\var{probability}}\approx 2^{-60.00}$
%and $\mbox{\var{target}}\approx 2^{256-60.00}$.

{\bf Remark 2.}
Initially in early 2009 the probability was only $2^{-32.0}$.
Back then it was 256 million times easier than today
to solve CISO puzzles.
Many early adopters of bitcoin
have made a lot of money. %actually only if kept bitcoins...
One of the well-known problems of bitcoin is the problem
of hoarding: a substantial proportion of bitcoins in circulation
is not used.

{\bf Remark 3.}
It is also widely believed that many bitcoins
have been lost because their owners did not think
it would ever be worth some serious money.
Even though all bitcoin data are public there is no way
to tell the difference between bitcoins which are saved
(and could be sold or exchanged later),
from those which have been lost.
Therefore is it not correct to believe that the monetary supply of bitcoins is fixed.
We can only say that it is upper bounded and limited by the existing production
and a cap of 21 million bitcoins to be ever produced.
However there is no way to know how many bitcoins are in {\em active} existence.
%%%%%IS IT TRUE??? I think they are not uniquely recognized? or are they? what if I move the same bitcoin around?
%However it is very easy to observe how many are actually in circulation
%because every single transaction is public.
%%%%%IS IT TRUE??? I think they are not uniquely recognized? or are they? what if I move the same bitcoin around?

%Similarly unless bitcoins are actually exchanged against real money,
%it is impossible for the fiscal authorities to know how many bitcoins exist
%in circulation.

\part{How To Speed Up The Bitcoin Mining Process}
\label{PartImprovedMinining}

%\newpage

\section{Is There A Better Way?}
\label{IsThereABetterWay}
\label{SectionIsThereABetterFifthGeneration}

In this part we look at a pure specialist question which pertains
to symmetric cryptography,
of whether there is a cryptographic ``shortcut'' attack:
simply a method of mining bitcoins faster than brute force,
or faster than the trivial method in which the SHA-256
hash function is a black box.
The answer is trivially yes, such a method trivially exists
and most developers of modern bitcoin miner hardware
have already applied various tricks which enhance the speed or/and decrease the cost.
However until now there was no public discussion of these questions
and it was not possible to see how far one can go in this direction.
In this paper we describe a series of more or less non-trivial optimizations
of the bitcoin mining process.
% and some ten more potential optimizations???????,
%which can be developed further.???????????????????????
These optimizations are quite important as considerable computing power
is already expended on our planet on bitcoin mining \cite{MiningMegaWattsEnvirDisaster}.

%\newpage

%\subsection{First Result}
%\label{SubSectionMethod01}

The question is what is the fastest possible method
for bitcoin mining, given the specific structure
of Fig. \ref{BitcoinCISOProblem}
and can we save some of the gates needed for this task.
The answer is yes we can.

In this paper we are the first to develop such techniques
openly and publish them.
We have invented these techniques independently from scratch and
to the best of our knowledge they are free of any intellectual property rights.
However we expect that ASIC designers have already done similar optimizations
and some of these techniques could have been patented.

{\bf Related Search:}
One could also try to solve this problem by
formal coding and ``a software algebraic attack'',
see \cite{desalg,FasterMiningSAT,RaddumDES}.

\newpage
\section{Bitcoin Mining: Past Present and Future}

\subsection{Four Generations of Bitcoin Miners}

Since 2009 bitcoin mining have gone through four major stages.
Speed of bitcoin operations is measured in GH/s or mega hashes per second,
as these operations are essentially
about computing the standard hash function SHA-256
many times.
No source gives a clear definition of H/s
as the speed of SHA-256 is variable and depends on data length.
We will go back to this question later.

\vskip-4pt
\vskip-4pt
\begin{enumerate}
\item
{\bf First generation - software mining using CPUs}.
Initially amateurs used to do these computations at home
with open-source software.
%Many bitcoins have been created with CPUs and this was initially profitable,
%however the difficulty of the task increased with time.

Various modern CPUs allow to achieve roughly between 1 and 5 MH/s per CPU core.
With this technology miners have been expending quantities of energy
to produce one Giga Hash per second.
For example we have computed that with Intel i5 processors
we would need some 50 4-core CPUs consuming 4000 Watts.
The power consumption is therefore 4000 W per GH/s.

\item
{\bf Second generation - software mining using GPUs}.
Graphic card CPUs have revolutionized bitcoin mininng
however they have NOT always achieved very important savings
compared to CPUs.
In some cases their electricity consumption
is not much lower than with CPUs.
%speedups of about 100 times.
%source: https://www.weusecoins.com/en/mining-guide
Other solutions are more efficient and allow one to mine with
a power consumption at least 10 times lower than with CPU mining see \cite{MiningHardwareComparisonWebPage}.
For example with Radeon 7790 we obtain about 0.33 GH/s with a power consumption of 70 watts.
This is about 210 W per GH/s.

\item
{\bf Third generation - hardware mining with FPGAs}.
Then miners have used FPGAs, not always achieving
much higher speeds on devices with comparable cost and size,
but decreasing the power consumption quite substantially,
up to 100 times in comparison to CPU mining.
For example ModMiner Quad based on a 45 nm FPGA requires
about 50 W per Gh/s.
%and therefore the cost of
%mining in the long run
%by a factor of roughly 5 times.

\item
{\bf Fourth generation - hardware mining with ASICs}.
Finally since mid-2013 %(and much earlier in development)
miners are moving towards using ASICs, dedicated hashing chips.
%Making of these have required literally millions of dollars of upfront investment which was paid...
This further decreases the cost of mining and in particular power consumption many times.
These devices can achieve as little as 0.35 W per Gh/s
(pre-order announcement from Bitmine.ch expected to ship in November 2013).
%A device from Butterfly Labs provides 50,000 MH/s and consumes maybe 200W of power
%in our estimation \cite{ButterflyLabs50GHMiner}.
%Currently all such devices in production have been sold several months before they have been manufactured.
\end{enumerate}
\vskip-4pt

As we can see,
the energy efficiency of bitcoin miners have improved by a factor of
nearly 10,000 since 2009.
Recent developments have driven amateurs out of business and require
them to invest thousands of dollars and purchase specialized hardware.
At the same tile new innovative business ventures
make money by selling increasingly sophisticated bitcoin mining devices.
At the moment of writing the key players in this business are the US company Butterfly Labs,
Swedish KNC miner, the Swiss company Bitmine.ch, their Russian competitor BitFury and few other.

\vskip-5pt
\vskip-5pt
\subsection{Electricity Consumption of Bitcoin Mining Operations}
\label{ElectrictyConsumption}
\vskip-5pt

There is abundant publicly available data about bitcoin mining.
In April 2013 it was estimated that bitcoin miners
already used about
%1 Giga Watt hour %
982 Megawatt hours every day,
enough to power about 30,000 U.S. homes or an equivalent
of 150,000 USD per day in electricity bills.
Still %current price of bitcoins
they would be able to make some 0.7 Millions of dollars in daily profits
\cite{MiningMegaWattsEnvirDisaster}.
At that time the hash rate was about 60 Tera Hashes/s.

At the moment of writing (22 October 2013)
the hash rate has attained 3000 Tera Hashes/s due to
a massive switch from GPU and FPGA mining to ASIC mining.
However the power consumption
have probably decreased
due to the fact that recent mining devices are more efficient,
see Section \ref{SpeedFinalResult1.8X}.

%This in addition to the cost of acquiring or building the mining devices.
Bitcoin mining is known to be a highly profitable business.
Some online tools for bitcoin profitability calculations
based on the price of electricity
%and excluding fixed capital expenditure spent on purchasing the computing devices
are available, cf. \cite{BitcoinProfitabilityCalculator1}.
%BitcoinProfitabilityCalculator1,
%these

%secret service command of people and what they do may be another explanation for hoarding

%More recently it has become an ecosystem in which many new technology and financial
%business ventures alike are trying to invent the future of payments,
%or the future of the financial industry.

\subsection{Towards a Fifth Generation of Miners}
%Is There Ever Going To Be a Fifth Generation?}
\vskip-5pt

We contend %In fact it is easy to see
that there will be further improvements in the basic technology.
In science, not everything can be improved.
Interestingly in business, %studies
we are accustomed to see that more or less every technology
which has some economic impact can be systematically improved every year.
This is for example is reflected in the famous Moore's law.
We see no reason why it should be otherwise
with basic algorithmic technology
behind bitcoin mining,
this independently from the question
of efficient hardware implementation of this technology.
Such improvements are inevitable.
In the long run, we believe that sooner or later there will be
substantially better technology for bitcoin mining,
would this be with quantum computers,
software algebraic attacks \cite{desalg,FasterMiningSAT,RaddumDES},
or a fundamentally different methodology than currently known.
In order to fix the ideas we call this
claim a {\em super optimistic assumption}.

The interesting aspect is that researchers who are able
to generate such improvements will be able to make a lot of money
by mining bitcoins and selling them at their market price,
or by licencing their algorithmic improvements to miners.
Moreover even a tiny energy efficiency improvement of 1 $\%$
could be profitable
as it will generate already thousands of dollars of tangible savings
on electricity bills.
%This situation is unprecedented in the history of cryptographic optimisation.
%Usually optimizations and improved implementations are hard to sell,
%as they are slow and costly to develop.
In this paper we show that such improvements are possible,
see Section \ref{SpeedFinalResult1.8X}.
However we do not claim that we are getting anywhere near
the fifth generation of bitcoin miners.
We have been only moderately successful in this task and therefore
our result are like generation 4.1. of bitcoin miners, a small improvement.
We offer our improvements free of charge and do not plan to patent them.

\newpage
\vskip-5pt
\vskip-5pt
\section{Description of the Problem}
\label{DescriptionOfHashing1}
\vskip-5pt

In this section we re-visit and expand our technical explanation
of the internals behind bitcoin mining fom Section \ref{CISOProblemInternals}.
We recall that we can see the problem of bitcoin mining
as a specific problem in symmetric cryptography which we
called ``CISO hash puzzle''.
It involves three applications of a block cipher.
We have already outlined this approach on Fig. \ref{BitcoinCISOProblem}
and now we explain it in all due details.
Our analysis follows the NIST specification of SHA-256
\cite{SHA256FIBS180}
and the inspection of the Bitcoin source code \cite{BitcoinMainSoftwareDistribution}.
We use vere similar notations and graphical conventions
as the leading experts of SHA-256 in the cryptographic literature,
see for example \cite{SHA256PreimageMatusiewicz,MatusiewiczModifiedSHA256}.
We start by recalling how the SHA-256 has function is constructed and then we show how
exactly it is used in bitcoin mining.

\begin{figure}[!h]
\centering
\begin{center}
\vskip-7pt
\vskip-7pt
\includegraphics*[width=3.72in,height=3.23in,bb=0pt 0pt 744pt 646pt]{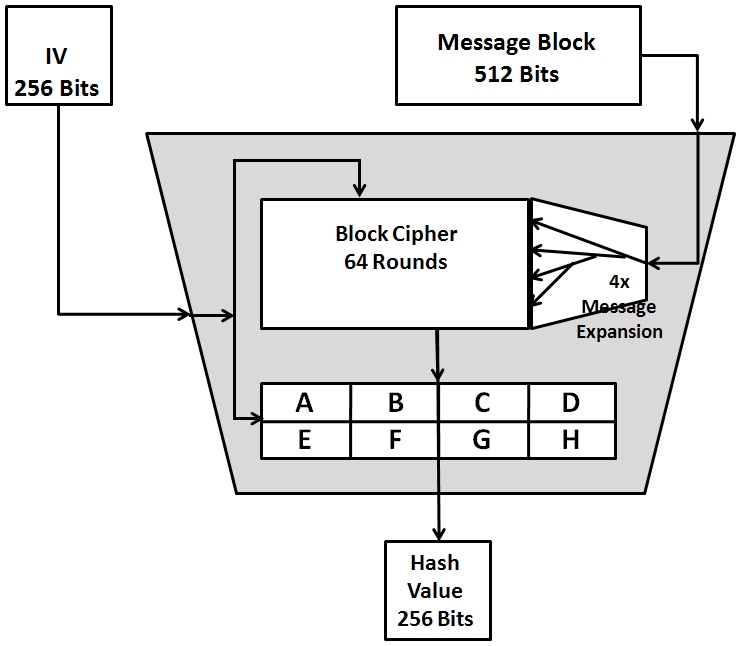}
\end{center}
\vskip-7pt
\vskip-7pt
\vskip-7pt
\caption{
One compression function in SHA-256.
It comprises a 256-bit block cipher with 64 rounds,
a key expansion mechanism from 512 to 2048 bits,
and a final set of eight 32-bit additions.
%said below
%The first $512=16\cdot 32$ message bits are simply copied.% in the first 16 rounds.
}
\label{Fig:OneCompression}
\end{figure}

SHA-256 is a hash function built from a block cipher following
the well-known Davies-Meyer construction in which
the input is at the end added to the output.
This construction is one of the known methods to transform a block cipher into a compression function.
A compression function is a building block of a hash function with a fixed input size.
It is typically equal to twice the output size.
In our case we have a compression function from 512 to 256 bits,
cf. Fig. \ref{Fig:OneCompression}.

%\newpage

%\vskip-2pt

%with three distinct applications of the block cipher which underlies SHA-256.
%Inside we find a block cipher.
The block size in this block cipher is 256 bits,
the key size is 512 bits which is expanded to 64 subkeys on 32 bits each
for each of 64 rounds of the cipher.
The first 16 subkeys for the first 16 rounds
are identical to the message and are copied in the same order
cf. \cite{SHA256FIBS180} and later Fig. \ref{MessageExpansion}.

In addition in order to hash a full message, SHA-256 applies a Merkle-Damgard
padding and length extension which makes it a secure hash function for messages of variable length.
In the pre-processing stage, we must append one binary 1 and many zeros to the message in
such a way that the resulting length is equal to 448 modulo 512, cf. \cite{SHA256FIBS180}.
Then we append the length of the message in bits as a 64-bit big-endian integer.
%append the bit '1' to the message
%append k bits '0', where k is the minimum number >= 0 such that the resulting message
%    length (modulo 512 in bits) is 448.
%append length of message (before pre-processing), in bits, as 64-bit big-endian integer

\begin{figure}[!h]
\centering
\begin{center}
\hskip-1pt
\hskip-1pt
\includegraphics*[width=4.72in,height=2.23in,bb=0pt 0pt 1162pt 478pt]{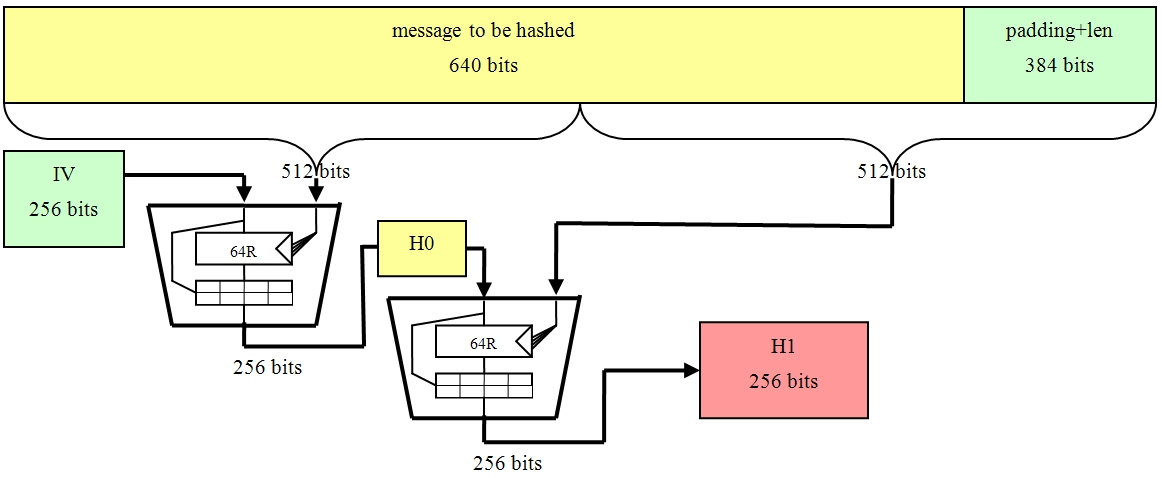}
\end{center}
\vskip-2pt
\caption{
The internals of SHA-256
when hashing a 640-bit message as used in
the first application of SHA-256 in bitcoin mining.
}
\label{Fig:FistSHA256640Bits}
\end{figure}
%\vskip-2pt

An interesting peculiarity in Bitcoin specification and source code is that
hashing with full SHA-256 is applied twice.
This may seem as excessive: one ``secure'' hash function should be sufficient.
It also makes our job of optimizing bitcoin
mining substantially more difficult.
In the first application of SHA-256 in Bitcoin mining the message
has a fixed length of 640 bits which requires
two applications of the compression function
as shown on Fig. \ref{Fig:FistSHA256640Bits}.
In the second application SHA-256 is applied to 256 bits.
Overall ``in theory'' we need three applications of the compression function
as already shown on Fig. \ref{BitcoinCISOProblem} which we also show on
a smaller-scale Fig. \ref{BitcoinCISOProblem2} below for convenience.

%....
%We arrived at three distinct applications of the block cipher which underlies SHA-256.
It may therefore seem that a bitcoin miner needs to compute
the compression function 3.0 times for each nonce and for each Merkle hash.
In the following sections we are going to work on reducing this figure
%down to about 1.86 on average.
down to about 1.89 on average.
Further details about inner mechanisms of SHA-256
will be provided later when we need them
cf. for example Section
\ref{Improvement4IncrementalRound3}.
%\cite{SHA256FIBS180}

\newpage

\subsection{Short Description of the CISO Hashing Problem}
\label{ShortDescriptionCISOWithSecondPictureAndLegend}
\label{RecallDefCISOProblem}
\vskip-5pt

We recall from Section \ref{FirstQuickDefCISOProblem} that new bitcoins can be created
when the miner succeeds to hash some data from the bitcoin
network together with a 32-bit random nonce and is able to obtain
a number on 256 bits which starts with a certain number of 60 or more zeros.

\begin{figure}[!here]
\centering
\begin{center}
\vskip-7pt
\vskip-7pt
\hskip-13pt
\hskip-13pt
\includegraphics*[width=4.60in,height=3.0in,bb=0pt 0pt 1158pt 666pt]{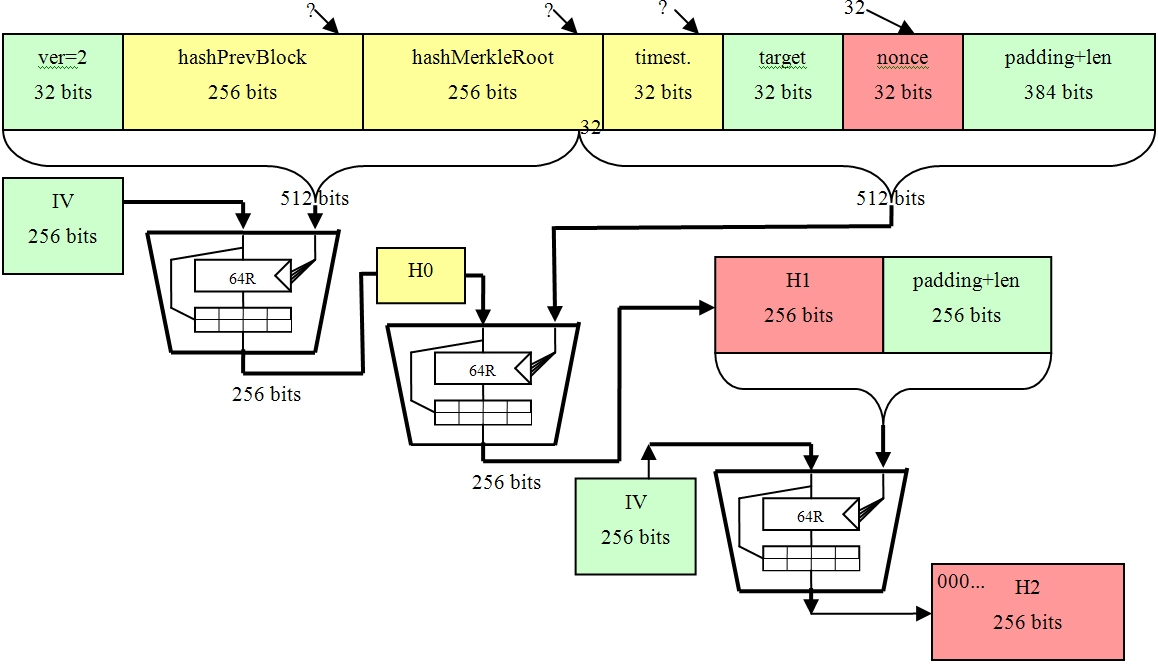}
\vskip-0pt
\vskip-0pt
\includegraphics*[width=4.60in,height=2.3in,bb=0pt 0pt 624pt 319pt]{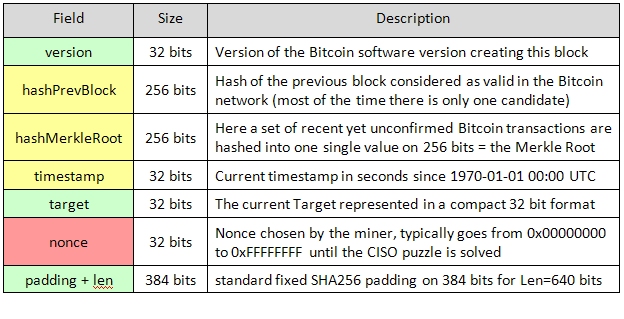}
\end{center}
\vskip-14pt
\vskip-14pt
\caption{
Our CISO problem seen
as three applications of the underlying block cipher
as in bitcoin mining.
}
\label{BitcoinCISOProblem2}
\vskip-3pt
\vskip-3pt
\end{figure}

We call it {\bf C}onstrained {\bf I}nput {\bf S}mall {\bf O}utput problem
or shortly the CISO problem.
On Fig. \ref{BitcoinCISOProblem2} we recall the key steps in this process.
The process needs to be iterated with different values of MerkleRoot and different 32-bit nonces
until a suitable ``CISO configuration'' is found in which
the output satisfies $H2 < \mbox{\var{target}}$
%seen as a bit integer being smaller than
as explained in Section \ref{TargetExplained}.

%\vskip-2pt

\newpage

\vskip-5pt
\vskip-5pt
\section{Our Optimizations}
\label{AllImprovements}
%\vskip-5pt
%\vskip-5pt
%\vskip-5pt
\subsection{Improvement 1: Remove First Compression Function}
\label{Improvement1RemoveFirstCompressionFunction}
\vskip-5pt

We can reduce the cost factor from 3.0 to 2.0 almost instantly
by making the following observation.
In the process of bitcoin mining the first compression function
does {\bf not} depend on the random nonce on 32 bits.
Therefore we can compute it once every $2^{32}$ nonces.
On average we need
$$2.0+\frac{1}{2^{32}}$$
compression functions.
The added factor is the amortized cost
of the first hash and can be neglected.

{\bf Important Remark.}
In more advanced bitcoin mining algorithms
the miner does {\bf not} have to compute the output for every nonce.
He can do it only for some well chosen nonces.
They may be chosen in such a way as in order to
obtain specific values which make the computation easier.
Moreover, some well chosen nonces could be generated
in some specific order in order to enable incremental computations.
In an incremental computation some computations could made easier
by reusing all the (known) internal values in one or several previous computations.
There is a lot of highly non-trivial optimizations which can be developed.
%In this paper we just do some very basic steps in this direction.
One simple example of incremental computation
will be given in Section \ref{Improvement4IncrementalRound3},
another in  Section \ref{Improvement8IncrementalTwoMoreAdditions19}.

\vskip-5pt
\vskip-5pt
\subsection{Improvement 2: Save 3 Rounds at the End}
\label{Improvement2RemoveRoundsEnd}
\vskip-5pt

We look at the computation of H2 on Fig. \ref{BitcoinCISOProblem2},
(the second computation of the hash function
and the third compression function).
A close  examination %of the hash function
reveals that in last rounds of the underlying block cipher
the two words on 32 bits in which we we want to have at least 60 zeros, after addition of a suitable constant,
are created at rounds 60 and 61 if we number from 0.
We basically want to force values created at rounds $t=60$ and 61 to two fixed constants which come from the SHA-256 IV constants,
and which would produce zeros at the output.
For this most of the time we just need to compute the first
61 rounds out of 64 and we can early reject most cases.
Only in $1/2^{32}$ of cases we need to compute 62 rounds in the third compression function.
Then only in some $1/2^{60}$ of cases where we have actually obtained at least 60 zeros,
we would need compute the full 64 rounds.

Thus overall one only needs to compute
the whole compression function an equivalent of
very roughly $1+61/64 \approx 1.95$ times on average.
Most of the time one only needs to compute $H1$ and 61 rounds of $H2$
to early reject the 32-bit value obtained which must be equal to
the IV constant.

\begin{figure}[!h]
\centering
\begin{center}
\vskip-3pt
\vskip-3pt
\hskip-1pt
\hskip-1pt
\includegraphics*[width=4.00in,height=0.72in,bb=0pt 0pt 600pt 104pt]{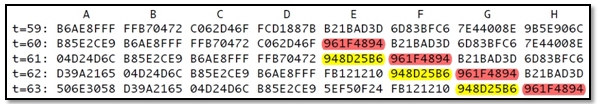}
\end{center}
\vskip-7pt
\vskip-7pt
\caption{
An example showing the last 5 Rounds of a SHA-256 computation.
}
\label{BitcoinCISOLast5RoundsExample}
\end{figure}
%\vskip-2pt

{\bf Remark 1.}
This figure is not exact and in fact it is slightly less.
This is because we can in fact save a higher fraction
of about 3/48 of the message expansion process when we stop our computation at 61 rounds.
This is due to the fact that message expansion is only computed in the last 48 rounds,
in the first 16 rounds the message is copied
cf. \cite{SHA256FIBS180} and later Fig. \ref{MessageExpansion}.
For the sake of simplicity we ignore the message expansion in our calculations.

{\bf Remark 2.}
We have carefully checked the ordering of words
by inspection of bitcoin source code \cite{BitcoinMainSoftwareDistribution} and by computer experiments.
An interesting question is what would happen if the bitcoin designers
have formatted the output of the hash function in the reversed order.
If they required that 60 bits are at 0 at the opposite end compared to the current formatting,
then it is possible to see that the miner would need to do more work:
63 out of 64 rounds in the last application of the compression function.
This would make mining more expensive and
would cancel a good proportion of %(about 2/3)
our %%%%%%%%%%50 MWh daily
savings.
%and we would obtain 1.9375 instead of 1.90625.
%overall with next step also...... this would cancel about 1/3 of our 50 MWh daily savings...

\vskip-5pt
\vskip-5pt
\subsection{Improvement 3: Reduce the Number of Rounds at the Beginning}
\label{Improvement3RemoveRoundsBegin}
\vskip-5pt

Now we look at the second computation of the hash function in the second compression function,
the computation of H1 on Fig. \ref{BitcoinCISOProblem2}.
Here we use the observation that in SHA-256 the key for the first 16 rounds
are exactly the 16 message blocks in the same order,
cf. \cite{SHA256FIBS180} and Fig. \ref{MessageExpansion}.
It is possible that in the second compression function on Fig. \ref{BitcoinCISOProblem2}
the nonce enters at round 3 (numbered from 0) and therefore in most cases
we just need to compute the last 61 out of 64 rounds of the block cipher.
The first three rounds are the same for every nonce and their (amortized) cost is nearly zero.

%also checked that diffusion is immediate on newly created values...
%%%%%no effect, no cost of copying... 2/8 of the state and propagates slowly in the next few rounds.
Putting together Improvement 2  and Improvement 3,
overall one only needs to compute
the whole compression function
slightly less than an equivalent of about
$2\times 61/64 = 1.90$ times.

\vskip-5pt
\vskip-5pt
\subsection{Improvement 4: Incremental Calculations in Round 3}
\label{Improvement4IncrementalRound3}
\vskip-5pt

\begin{figure}[!h]
\centering
\begin{center}
\vskip-9pt
\vskip-9pt
\hskip-1pt
\hskip-1pt
\includegraphics*[width=4.50in,height=3.8in,bb=0pt 0pt 580pt 391pt]{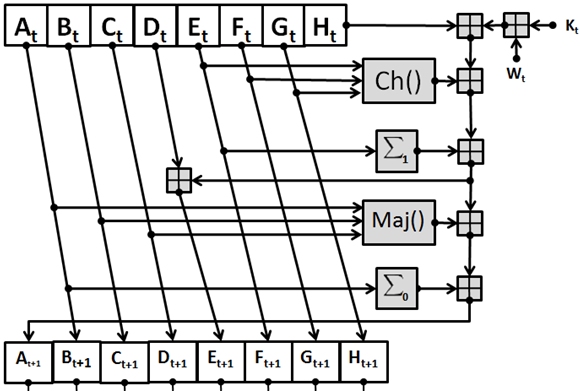}
\end{center}
\vskip-7pt
\vskip-7pt
\caption{
One round of the block cipher inside SHA-256
}
\label{BitcoinOneRoundSHA256}
\end{figure}

\begin{figure}[!h]
\centering
\begin{center}
\vskip-3pt
\vskip-3pt
\hskip-1pt
\hskip-1pt
\includegraphics*[width=4.66in,height=4.72in,bb=0pt 0pt 466pt 472pt]{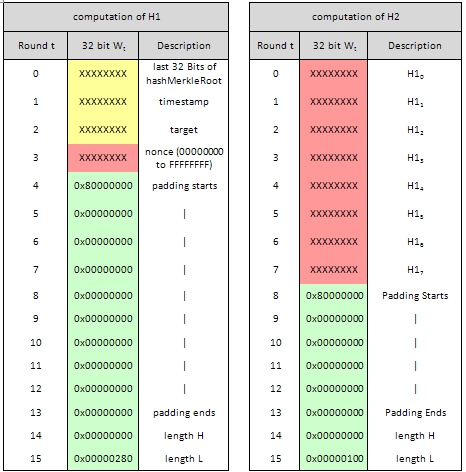}
\end{center}
\vskip-7pt
\vskip-7pt
\caption{
Key in the first 16 rounds out of 64 in each computation and their provenance.
}
\label{ZeroAndConstantKeysFirst16Rounds}
\end{figure}

This improvement requires us to delve more deeply
into the structure of the block cipher inside SHA-256.
We recall that the state of the cipher after round 2 is constant
and does not yet depend on the value od the nonce.
The 32-bit nonce will be precisely copied
to become the session key for the round 3 of encryption.
On Fig. \ref{BitcoinOneRoundSHA256} we show the circuit %diagram
for one round of encryption where at round 3 the nonce enters as
$W_3=\mbox{\var{nonce}}$ as shown on later Fig. \ref{ZeroAndConstantKeysFirst16Rounds}.
Here $\boxplus$ denotes one addition on 32 bits.

%looks as follows:

\newpage

\vskip-5pt
\vskip-5pt

Here $W_t$ is the key derived from the message and $K_t$ is a certain constant \cite{SHA256FIBS180}.
For $t=3$ we have $W_3=\mbox{\var{nonce}}$. % as shown on Fig. \ref{ZeroAndConstantKeysFirst16Rounds}.
Now it is obvious that the whole round 3 %with 7 additions and other operations
can be computed essentially for free in the incremental way.
We just need two 32-bit increments
instead of one whole round which is about 7 additions
and 4 other 32-bit operations.
%which for simplicity we assume to have a similar cost.
%Overall we save about 90 $\%$ of the computation of the round 3.
Each time we increment the nonce we simply need to increment two values
(in columns A and E)
at the output of round 3, which is shown on Fig.
\ref{BitcoinOneRoundSHAIncrementalComputationR3} below.

\begin{figure}[!h]
\centering
\begin{center}
\vskip-3pt
\vskip-3pt
\hskip-1pt
\hskip-1pt
\includegraphics*[width=4.60in,height=0.9in,bb=0pt 0pt 1138pt 228pt]{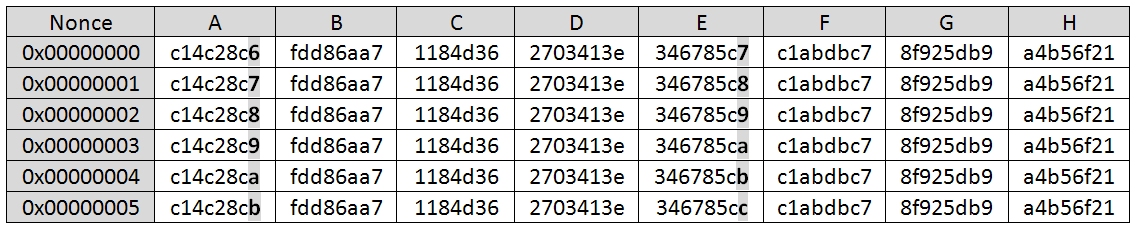}
\end{center}
\vskip-7pt
\vskip-7pt
\caption{
Example of incremental direct computation of the state after round 3.
}
\label{BitcoinOneRoundSHAIncrementalComputationR3}
\end{figure}
%\vskip-2pt

Thus we have saved one more round in each
of our computations.
% of the whole circuit computed for each nonce.

\newpage
\vskip-5pt
\vskip-5pt
\subsection{Improvement 5: Exploiting Zeros and Constants in the Key}
\label{Improvement5ZerosConstantsWt}
\vskip-5pt

The next improvement comes from the fact that the key
in the first 16 rounds of the block cipher is an exact copy of the
message. Many parts of this key are constants.
Many are actually always equal to zero.
This allows one to save a lot of additions in the computation of SHA-256.

%\vskip-2pt

Let
$KW_t=K_t\boxplus W_t$.
Overall we see that we can save 18 additions:
16 additions have a constant equal equal to zero,
and $KW_t=K_t$
and we have 2 more additions
with 0x80000000
which can be replaced by flipping one bit,
the cost of which is very small (in hardware).
%when compared to the cost of one addition.

\newpage
\vskip-5pt
\vskip-5pt
\subsection{Improvement 6: Saving Two More Additions
 with Hard Coding}
\label{Improvement6HardCodingFiveAdditions}
\vskip-5pt

It is easy to see that 2 more additions can be saved.
Looking at Fig \ref{ZeroAndConstantKeysFirst16Rounds} we should not count
the three first constants on the left in yellow which are identical
for all the $2^{32}$ different nonces.
This is because this saving was already done
in Section \ref{Improvement3RemoveRoundsBegin}.
However we have two additional constants in the last line in green.
Then in these two last rounds, one in each computation,
and because the $K_t$ are constants and do not depend on the message being hashed,
we can pre-compute the constants $KW_t=K_t\boxplus W_t$ on
32 bits which saves us 2 additions such as in the upper right corner
of Fig. \ref{BitcoinOneRoundSHA256}.

%We have a total of 6+6+1 rounds and 18+2 additions which have been saved until now.
%compare GE

{\bf Remark:}
In Section \ref{Improvement5ZerosConstantsWt}
above and here in Section \ref{Improvement6HardCodingFiveAdditions}
we saved 18+2 additions.
Moreover the output of these two additions is
a constant in 18+2 cases saving not one but
first two additions
in the upper right corner
of Fig. \ref{BitcoinOneRoundSHA256}.
However these savings are illusory,
because we can {\bf save many more additions} by another method.
For example later on Fig. \ref{BitcoinOneRoundSHA256WithCSA}
we are going to show that one only needs essentially 2 additions
in order to implement the whole round function of SHA256,
this instead of 6+1 full adders
as on Fig. \ref{BitcoinOneRoundSHA256}.

\vskip-5pt
\vskip-5pt
\subsection{Known Results About SHA256 in Hardware/ASIC}
\label{ImprovementWhatIsAlreadyKnown}
\vskip-5pt

The goal of this paper is NOT to describe the best possible ASIC implementation of SHA256
and what is the best compromise between the circuit area/propagation time,
%Several very good implementations of SHA256
%are already described in the literature,
see \cite{ChavesSHA256Hardware,DaddaASIC_SHA2,DaddaASIC_SHA2High,RechbergerCompactSHA256,KnezevicHardwarePhD,IterationThroughputSHA256Verbauwhede,DaddaQuasiPipeSHA,CyprusSHA256Implementation,CyprusSHA256Implementation2,SHA256GEPercentages,SklavosHardSHA256,TillichHardwareAllSHA3}. We work on a slightly different problem:
we show that the problem goes beyond any state of the art SHA256 circuit, and has its own specificity.
%and we don't go as far as a full 2x or 4x unrolled implementation which optimizes the propagation time and carefully uses CSA blocks instead of additions with tricks to get more area
%but higher speed nevertheless... Etc.
%We don't take these things to their ultimate stage of a pipelined careful design.
The novelty in our paper is adaptation to mining specifically,
such as most hashed message blocks are constants,
and 6 block cipher rounds do NOT need to be computed at all, incremental techniques etc.

\newpage
\vskip-5pt
\vskip-5pt
\subsection{Improvement X - Saving LOTS More Additions With Delayed Carry Propagation}
\label{ImprovementSabingLOTSWithDelayedCarryPropagation}
\vskip-5pt

We are going to use
Carry Save Adders (CSA)
in order to delay the propagation of carries
and save a lot of circuit area.\footnote{
We call it Improvement X and we do not give it a number in this paper 
as this improvement also concerns full SHA256 implementations in ASIC 
and has already been applied by most ASIC designers. 
In this paper we focus on the difference between fully functional 
general-purpose ASIC and a specific solution for bitcoin mining. 
}
The main idea
which is attributed to John von Neumann.
is to propagate
the carries only locally
delaying a complete propagation
to the very end.
This allows a dramatic reduction in the cost of
implementing multiple additions:
three or more additions
cost do NOT cost much more than
%essentially as much as
one single addition.

More precisely Carry Save Adders (CSA) allow
to add $n$ numbers for any $n\geq 3$
and to form two numbers %ps and sc
which need to be added to obtain the final result.
This is obtained by a successive transformation of
3 numbers into 2 numbers with a Carry Save Adder (CSA)
which has a very low cost
and a final addition of 2 numbers.

\vskip-5pt
\vskip-5pt
\begin{defi}[Carry Save Adder (CSA)]
A Carry Save Adder takes 3 integers
$a,b,c$ on $k$ bits written in binary
and outputs two numbers
ps (partial sum) and
sc (shift-carry) as follows:
%which need to be added to obtain the final result.
\vskip-5pt
\vskip-5pt
$$
ps_i = a_i \oplus b_i \oplus c_i
$$
\vskip-5pt
\vskip-5pt
$$
sc_{i+1} = a_i b_i \vee a_i c_i \vee b_i c_i
$$
\vskip-5pt
\end{defi}
\vskip-5pt

\begin{figure}[!h]
\centering
\begin{center}
\vskip-9pt
\vskip-9pt
\hskip-1pt
\hskip-1pt
\includegraphics*[width=3.6in,height=2.0in,bb=0pt 0pt 360pt 200pt]{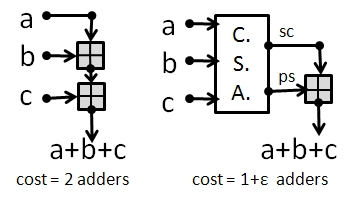}
\end{center}
\vskip-7pt
\vskip-7pt
\caption{
How to replace two adders by one adder
}
\label{ReplaceCSAPrinciple}
\end{figure}
\vskip-5pt
\vskip-5pt

In the case of addition modulo $2^{k}$
there is a slight simplification
as the most significant digits can be discarded
but the result remains essentially the same.

Overall the application of Carry Save Adders (CSA)
allows us to implement each round of SHA256
with only two additions
%REMOVED, I don't want to enumerate all the papers here again... would need to check if all use CSA
%\footnote{Application of
%Carry Save Adders (CSA) in implementation of SHA256
%has been studied before by many authors,
%see for example \cite{CyprusSHA256Implementation,CyprusSHA256Implementation2}.
%In this paper however we do in the context of bitcoin mining,
%which allows additional simplifications.
%Most of the time we do not need to perform the full SHA256 computation
%and many pieces of hashed data are constants.
%}
:

\newpage

\begin{figure}[!h]
\centering
\begin{center}
\vskip-9pt
\vskip-9pt
\hskip-1pt
\hskip-1pt
\includegraphics*[width=4.50in,height=3.8in,bb=0pt 0pt 700pt 440pt]{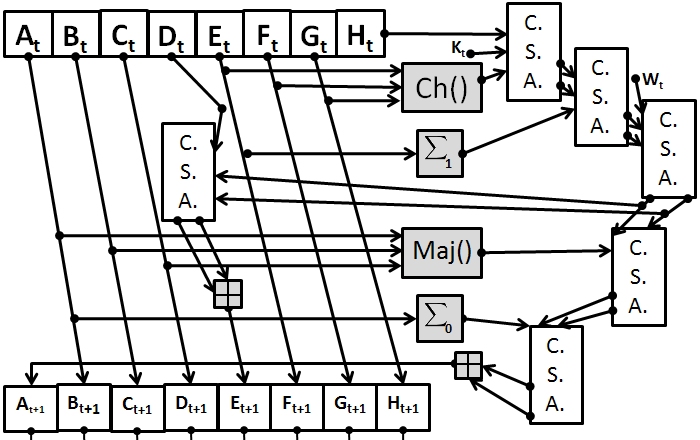}
\end{center}
\vskip-7pt
\vskip-7pt
\caption{
How to compute one round of SHA-256 with just two full adders
}
\label{BitcoinOneRoundSHA256WithCSA}
\end{figure}
\vskip-5pt
\vskip-5pt

Now we are also going to look beyond the 64 rounds of SHA256 seen as a block cipher.
What remains is the key expansion which expands the message to be hashed into
the 64x32 bit keys $W_t$.

\vskip-5pt
\vskip-5pt
\subsection{Short Description of Message Expansion in SHA-256}
\label{ShortDescriptionOfMessageExpansion}
\vskip-5pt

Before we can propose additional optimizations,
we need to explain how the message expansion works
in the NIST specification of SHA-256 \cite{SHA256FIBS180}.
%It works as follows.
We refer to \cite{SHA256FIBS180}
for definitions of $\sigma_0$ and $\sigma_1$.

\vskip-6pt
\vskip-6pt
\begin{figure}[!h]
\centering
\begin{center}
\vskip-3pt
\vskip-3pt
\hskip-1pt
\hskip-1pt
\includegraphics*[width=2.7in,height=1.32in,bb=0pt 0pt 540pt 264pt]{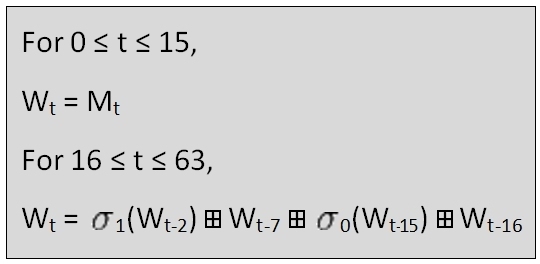}
\end{center}
\vskip-7pt
\vskip-7pt
\caption{
The message scheduler expanding a 512 message block into
a 2048-bit key for the SHA-256 block cipher.
%Here $\boxplus $ denotes the addition on 32 bits.
}
\label{MessageExpansion}
\end{figure}
%\vskip-2pt

%TBD re-write the above Fig with \boxplus:
%$$
%W_{t} = \sigma_1 (W_{t-2}) \boxplus W_{t-7} + \sigma_0 (W_{t-15}) \boxplus W_{t-16}
%$$

\vskip-5pt
\vskip-5pt
\subsection{Improvement 7: Saving Two More Additions}
\label{Improvement7HardCodingThreeMoreAdditions}
\vskip-5pt

We consider the computation of H1.
It is possible to see that the first two non-trivial keys
$W_{16}$ and $W_{17}$ %and $W_{18}$
are also constants and do not yet depend on the nonce.
This is because following Fig. \ref{MessageExpansion}  we have:

$$
W_{16} = \sigma_1 (W_{14}) \boxplus W_{9} \boxplus \sigma_0 (W_{1}) \boxplus W_{0}
$$

and

$$
W_{17} = \sigma_1 (W_{15}) \boxplus W_{10} \boxplus \sigma_0 (W_{2}) \boxplus W_{1}
$$

We see that these two values depend only on values which are constants on Fig.
\ref{ZeroAndConstantKeysFirst16Rounds}.
Therefore they can be pre-computed and as in Improvement 6 we can replace
the traditional hard-coded constants $K_t$ by
new hard-coded constants
$KW_t=W_t\boxplus K_t$
for the whole range of $2^{32}$ nonces.
Thus we save two more additions.

%We have a total of 6+6+1 rounds and 18+2+2 additions which have been saved until now.
%compare GE

\vskip-5pt
\vskip-5pt
\subsection{Improvement 8: One More Additions Saved by Incremental Computation}
\label{Improvement8IncrementalTwoMoreAdditions19}
\vskip-5pt

Now still in the case of H1 we have:

$$
W_{19} = \sigma_1 (W_{17}) \boxplus W_{11} \boxplus \sigma_0 (W_{4}) \boxplus W_{3}.
$$

Here $W_{3}$ is the nonce which is incremented by 1.
Other elements do not depend on the nonce and change at a much slower rate.
We can thus save one more addition and simply increment $W_{19}$ directly for each consecutive nonce.
%We save one more addition.

%old outdated: We have a total of 6+6+1 rounds and 18+2+2+1 additions which have been saved.
%compare GE

{\bf Remark:}
We also have:

$$
W_{18} = \sigma_1 (W_{16}) \boxplus W_{11} \boxplus \sigma_0 (W_{3}) \boxplus W_{2}
$$

Here the incremental computation is also possible though less efficient.
For the sake of simplicity we ignore these additional savings.

\vskip-5pt
\vskip-5pt
\subsection{Improvement X2: More Additions Saved By Delayed Carry Propagation}
\label{Improvement8IncrementalCSAInKeyScheduling}
\vskip-5pt

We see in the two above sections that three values $W_t$ require zero additions
and cost essentially nothing.
There remains 48-3 values $W_t$ to be computed.
If we use again Carry Save Adders (CSA)
we see that in each of the 45 cases
we need just one full adder instead of 3 full adders.

%\newpage

\newpage

\vskip-5pt
\vskip-5pt
\section{Our Overall Result on the Speed of Hashing}
\label{SpeedFinalResult1.8X}
\vskip-5pt

We have obtained the following result:
%We have the following result:

\vskip-5pt
\vskip-5pt
\begin{fact}[Hash Speed]
\label{HashSpeed1.8X}
The amortized average cost of trying
one output H2
to see if it is likely to have 60 or more
leading zeros
is at most about
%1.86 computations
1.89 computations
of the compression function
of SHA-256 instead of 3.0,
%which represents an improvement by $38 \%$.
which represents an improvement by $39 \%$.
%or equivalently the computation of the
This is an {\bf additional} improvement relative to already optimized ASIC implementations
of full SHA256
and we do NOT count substantial additional savings
which are obtained from Carry Save Adders,
pipelining and other well-known techniques.
\end{fact}
\vskip-5pt

\noindent\emph{Justification:}
We have saved about 7 rounds and many additions.
However known ASIC implementations also save many additions
and actually the designs which achieve the lowest possible area are not necessarily the fastest.
Therefore we are just going to estimate the RELATIVE savings
w.r.t. best standard ASIC implementations of full SHA256
such as in \cite{ChavesSHA256Hardware,DaddaASIC_SHA2,DaddaASIC_SHA2High,RechbergerCompactSHA256,KnezevicHardwarePhD,IterationThroughputSHA256Verbauwhede,DaddaQuasiPipeSHA,CyprusSHA256Implementation,CyprusSHA256Implementation2,SHA256GEPercentages,SklavosHardSHA256,TillichHardwareAllSHA3}. %We estimate
%(following one paper on hardware implementation of SHA-256)
%that one addition costs about 200 gates
%compared to maybe about 2500 gates for one full round of SHA-256
%cf. Table 3 in \cite{SHA256GEPercentages}.
%Thus as a rule of thumb we can consider that saving one addition
%is equivalent of saving $200/2500=0.08$ rounds.
Thus overall cost minus savings are
equivalent to a total of

$$
\frac{64+64}{64}-\frac{7}{64} = 2-\frac{7}{64} \approx 1.89 \mbox{~~compression functions}
%removed on 07042014   +23\cdot 0.08
$$

%This 1.86 compression functions is equivalent to saving of $38 \%$
This 1.89 compression functions is equivalent to saving of $37 \%$
compared to the initial cost of 3.0 compression functions
as per Fig. \ref{BitcoinCISOProblem2}.
It also shows how much can be gained in bitcoin mining
compared to using an optimized SHA256 ASIC implementation three times.

{\bf Remark 1.}
Fact \ref{HashSpeed1.8X} might seem a bit trivial.
In fact it allows miners to {\bf save up to about 50 Megawatt hours every day},
or some 75,000 USD per day in electricity bills,
if we assume the reported daily consumption of bitcoin mining in April 2013
\cite{MiningMegaWattsEnvirDisaster}.
This would be some 28 million dollars per year in savings.

{\bf Remark 2.}
However at the moment of writing (22 Oct 2013) things are expected to be different.
The aggregated hash rate in the network has increased
to 3000 Tera hash, roughly a 50 times increase from April 2013.
However we don't expect to consume as much as 1 GWh per day consumed announced in
\cite{MiningMegaWattsEnvirDisaster} because it appears that the power consumption has
decreased and not increased in the recent months,
this due to more efficient ASIC miners and to progressive
phasing out of inefficient miners which are no longer profitable.
If we assume that the average power consumption today is 3.2 W per Terahash
as with ButterFly Labs devices \cite{ButterflyLabs4GHASIC},
it gives 9.6 Mega Watts of estimated consumption for miners
or some 230 Megawatt hours per day. this is about 4 times less
than in \cite{MiningMegaWattsEnvirDisaster}.
If these estimations are correct
then the savings as of today would be only roughly a quarter of 75,000
daily saving of April 2013.
This will nevertheless amount to some 7 millions of dollars saved each year.
%hash power graph does not appear here

{\bf Remark 3.}
We ignore if recent mining devices already use Fact \ref{HashSpeed1.8X}
which developers could have discovered independently.
%Otherwise one can really modify their code and save 50 MWh per day.
We have noticed that the Swedish company KNCminer have explained
in some Internet forums
that they do indeed use some optimizations \cite{KNCMinerspecs}
which allow them to save about 30 $\%$ of the cost of mining.
%% full version maybe %%This is expected to be based on some advanced algorithms provided by \url{orsoc.se}.
%official on their web page: KNC miner is a joint venture between two strong companies, ORSoC AB and Kennemar & Cole AB
%advertises sth like this, 30% saving, spotted on 15052013

{\bf Remark 4.}
%Any further improvement in this 1.86 factor is going
Any further improvement in this 1.89 factor is going
to allow bitcoin miners to save tens of thousands of dollars per day on electricity.
It is also likely to influence the market price of bitcoins in the future.

{\bf Remark 5.}
Our problem is essentially the same as a brute force attack on a block cipher.
The same computation is done a very large number of times,
%Our problem is essentially about computing the same block cipher function
%a large number of times,
yet cheaper, maybe just a small factor cheaper.
%%easier here, I can CHOOSE A SUBSET OF KEY SPACE
%%which maximizes the weak keys percentage or maximizes incremental computations.
It is not correct to believe that
block ciphers are well understood in cryptography.
On the contrary,
it appears that for more or less any block cipher
there may exist an attack which will be  just slightly faster than brute force,
see \cite{AttackOnAnyBlockCipherGOST}. %,gostac}.
An efficient low-data software algebraic attack
could also be a solution to this problem, cf. \cite{desalg,FasterMiningSAT,RaddumDES}.
We expect that the future will bring many new developments in this space.

\newpage

\part{The Unreasonable Long-Term Property}
%\part{The Unreasonable Long-Term Properties}
%Towards A Reform of Bitcoin}
%\label{PartBitcoinNeedsReform}

\vskip-2pt
\vskip-2pt
%\section{A Reform? Is There A Need to Reform Bitcoin?}
%\section{What Is So Unreasonable?}
\section{The Unreasonable Artificial 4 Year Cycle}
\vskip-2pt

%% full version %%The title may appear very surprising.
Is there anything wrong with bitcoin at all?
%Why would anyone would like to reform bitcoin.
In the following sections we are going to argue that bitcoin has
at least one property which is truly unreasonable and
which needs to be changed in the near future.
%
%In sections which follow %this Part \ref{PartBitcoinNeedsReform}
In fact we are going to discuss something which has not yet been observed.
A predicted cycle of 4 years in bitcoin markets and data.
However bitcoin has been in existence for just over 4 years
and market data have been heavily distorted and blurred
by tremendous growth of the number of bitcoins in circulation and large capital inflows.
Under these circumstances our obscure 4-year property has not attracted a lot of attention.
It is however an undeniable fundamental
and built-in feature of the
current bitcoin virtual currency system
and it cannot possibly be ignored.
%% full version maybe %% and it has very serious consequences.

\vskip-2pt
\vskip-2pt
\subsection{The Artificial 4 Year Reward Cycle}
%\section{The Big Cycle}
\label{TheBigCycle}
\vskip-2pt

We recall the mechanism known as Block Reward Halving
which exists in all current bitcoin software \cite{BitcoinMainSoftwareDistribution,OpenSourceMiningApps}.
Every 4 years the amount of bitcoins awarded to a successul miner will be divided by 2.
It is 25 BTC as of 2013.

The origins of this property are obscure.
It was NOT proposed in the original paper of 2008 cf. \cite{SatoshiPaper}
which simply says that {\em any needed rules and incentives can be enforced}.
It simply is {\bf a fact hard coded in the current software}, cf. \cite{BitcoinMainSoftwareDistribution,OpenSourceMiningApps}.
%and its numerous clones and front-ends.
This mechanism is very closely related to the fact that the monetary supply of bitcoins is fixed to 21 million.
In fact this is how the 21 million cap is implemented.
The reward halving and 21 million cap are two faces of the same property.
In this paper we describe it only briefly,
we refer to \cite{BitcoinControlledMonetarySupply,Bitcoin21MillionCap}
for a more detailed explanation and discussions concerning the reasons why
this peculiar property is as it is.
%% full version %%following Section \ref{RewardCycle4Years}.
It may seem that this property is a sort of software bug,
however contrary a the software bug this property is very frequently applauded and praised.
It appears that we are the first to seriously criticize this property.

To summarize we have a 4 year reward cycle in the bitcoin digital currency as we know it.
Here is how exactly this mechanism works.
Initially prior to November 2012 all new CISO hashes were rewarded with 50 BTC.
Currently it is at 25 BTC for all blocks starting at 210,000
This reward price is going to stay stable until roughly end of 2016,
and then it will drop to 12.5 BTC for another period of 4 approximately years.
Then in each period of 210,000 blocks the reward is going to be halved again.
%At regular intervals of roughly about 4 years this
%price reward is halved.

More precisely every 210,000 blocks at every block which is an exact multiple of 210,000
the reward is halved for the next 210,000 blocks.
It is a sudden abrupt change which takes place approximately every
4 years (but not exactly) depending on the actual speed with which the block have been generated.
We should note that 210,000 is exactly 1 $\%$ of the 21 million
and we have an infinite geometric progression with a finite sum.
Our mechanism can be described by the following formula.
Let $t=210000*f$, the reward for any period of time $t\geq 0$ is:

\vskip-7pt
\vskip-7pt
$$
\mbox{\var{reward}}_{t\in [210000\cdot f\ldots 210000\cdot f+209999]}
~~=~~
%\mathbf{
50\cdot 2^{-f}
%}
\mbox{~BTC~~~for any~~} f\geq 0.
$$
\vskip-5pt

Then it is easy to see that this mechanism is precisely how the cap of 21 million of bitcoins to be ever generated is enforced.
More precisely if we count all the bitcoins ever produced with the current mechanism we obtain:

\vskip-7pt
\vskip-7pt
$$
%\mbox{\var{reward}}_{t\in [210000\cdot f\ldots 210000\cdot f+209999]}
%\mathbf{
\sum_{f=0}^{\infty}
210,000\cdot 50\cdot 2^{-f}
~=~
210,000\cdot 50\cdot (1+1/2+1/4+\ldots)
~=~
21,000,000
$$
\vskip-5pt

{\bf Summary.}
The key point is that with the current mechanism,
at fixed moments in time
it suddenly becomes twice more costly to mine bitcoins.
These {\bf sudden jumps} occur every 4 years
and are bound to have very serious financial
%and social
consequences.
It suddenly makes miners stop mining and switch their devices off.
Overnight. This must lead to serious
perturbations in the market.

The fact that this depreciation of the work of miners
happens by sudden jumps every 4 years is very surprising.
It is bound to have some serious consequences in all bitcoin-related markets. %from the point of view of economics and finance
%%%%%%%%%%some said below
%%%If this cycle is very strong or if just the ripple which happens
%%%every 4 years is very strong, it is possible to believe that this property
%%%makes that can hardly claim to be a liquid and stable financial asset,
%%%and this somewhat discredits bitcoin as it is today as a ``currency''
%%%or at least it seriously limits the applications of this currency.

%\newpage
\vskip-2pt
\vskip-2pt
\subsection{Artificial Cycle and Instability By Design}
%\subsection{Business Cycle - What is Really Wrong With Bitcoin As A Currency
%and What Makes It A Perfect High Tech Industry}
\label{KillerBusinessCycleArgument}
\vskip-2pt

%There is absolutely no denial that
With current bitcoin software \cite{BitcoinMainSoftwareDistribution,OpenSourceMiningApps},
at certain moments in time the reward for mining is divided by two in one single step.
This is NOT compensated by the fact the the difficulty of mining increases all the time.
It just adds sudden adjustments every 4 years to a difficulty curve which typically
goes systematically up due to a steady increase in the production of new hashes.
%if super highly profitable with margins>50%, no slump, but sooner or later profits<50%
We predict that in the future the \var{difficulty} curve will have a discontinuity at the moment
of the 4-year halving. Until now it has not happened because apparently
only very small percentage of active mining devices were switched off on 29 November 2012.

%It is not likely either to be compensated by capital inflows.
%BEWARE: the production of BTC is constant! also at this moment!
%BEWARE: the production of BTC is constant! also at this moment!
%BEWARE: the production of BTC is constant! also at this moment!
%It is unreasonable to think that on one single day there will be capital inflows to
%compensate
%and considerably inflate the price of bitcoin.
%On the contrary. There will be an inevitable slump in production of new bitcoins
%as on a single day many older mining devices will stop being profitable.
Inevitably, on one day many devices will stop being profitable and many people may lose their interest in bitcoin.
We are talking about the human factor.
Investors may decide that they are no longer going to give lots of money to a high-tech industry
which has just decreased production of hashes per second
and is binning many mining machines
at a massive scale due to a strange
rule which has no justification and could easily be modified.
They might move their money elsewhere and invest in another cryptocurrency.
Overall we expect that a sudden slump in profitability of mining
is likely to provoke some sort of {\bf a much larger ripple} in the bitcoin markets,
potentially lasting up to 4 full years.

To summarize we claim that the current bitcoin reward rule has important consequences.
It creates an artificial economical cycle
for the whole the industry of bitcoin mining,
for investors, and for traders who trade bitcoins.
There will be large capital inflows and outflows,
there will be privileged moments to invest money
and make profits.
Likewise it is extremely likely that there will be periods of time
of excessive production of SHA-256 hashes which will no longer be profitable.
Depending on the market price offered for bitcoin an over-production might force
the miner of less profitable devices to switch them off earlier than
at the boundary privileged moment where the reward is halved.
%when profitability will be negative and money will be lost.
%
%It is also quite possible that because many machines will be switched off on one single day,
%just before that day they will be over-production of hashes which will make people switch their machines earlier.

\vskip-2pt
\vskip-2pt
\subsection{The Inextricable Dilemma of Reforming How Bitcoin Works}
\vskip-2pt

Interestingly this cyclical property is easy to fix,
bitcoin technical authorities and developers
and stake holders can agree
to patch the bitcoin protocol,
and to smooth the thresholds of delivery of new bitcoins
to bitcoin miners.

However they will hesitate a lot
%This is a very difficult dilemma.
and will be faced with a certain dilemma:

\vskip-6pt
\vskip-6pt
\begin{enumerate}
\item
Either we
will be criticized by the financial press and media
%and sooner or later it will become clear that
that
{\bf bitcoin is not exactly as stable %and trustworthy
as government-issued currencies}
%from the pure financial point of view}.
%it lacks liquidity, stability
%and should never ever be used as such.
and that it has some
truly unreasonable cyclic properties.

\item
Or we will fix this problem.
Technically it is extremely easy.
It just requires a majority of people to agree.
Then suddenly bitcoin could become more stable
and therefore more like a currency.
%even though another strong cycle due to the development of new generations
%of miners
\end{enumerate}
\vskip-4pt

Now if we change the way in which bitcoins are awarded,
any decision made will have very serious consequences and
it would be extremely difficult to change again.

\vskip-2pt
\vskip-2pt
\subsection{Can The Reward Halving Cycle Be Compensated By Fees?}
\vskip-2pt
In the current bitcoin system there is no obligation whatsoever to include any fees in transactions.
A transaction fee will be decided by the sender of money
decided at the moment of making a transfer from one bitcoin address to another
and it will be an incentive for the miner to include the current transaction,
and the miner is able to collect hundreds of transaction fees for each block generated with no effort.
In practice current bitcoin software applies fees to most transactions
however the users are also able to override this behavior and sent transactions without fees
knowing that they will take longer to be confirmed.
%% full version %%
%% full version %%
%% full version %%{\bf Remark:}

It is unthinkable that fees would increase in a very substantial way overnight.
Therefore the income from fees is probably not going to compensate
the cyclic property we have described.
However at the very moment when the profitability of mining goes down,
miners used to a certain level of income will try to increase their income from fees.
This can for example by done by innovating in some way.
For instance miners might promote some method to achieve greater anonymity on the bitcoin network
which will split the usual transactions into many smaller transactions
and use large number of freshly generated addresses.
An increased number of transactions processed in each block would allow
to claim more fees, and bitcoin users are likely to accept to spend more on fees
as a price to pay for better anonymity.
Overall we believe that yes, the sudden slump in the reward is going to increase the income from fess,
but this is not going to happen overnight. On the contrary, it is one of these economic mechanisms
which will extend a predicted sudden slump at the boundaries of the cycle
to a much larger and much ``slower moving'' economic cycle lasting up to the full 4 years.

%\newpage

\vskip-5pt
\vskip-5pt
\subsection{A Proposed Fix}
\vskip-5pt

We propose that the block reward should be decremented
with a substantially higher frequency
than after a whole very long period of 210000 blocks.
%We take advantage of the fact that 1
%BTC is divisible up to 8 decimal places.
We are quite conservative and propose to leave sufficient
time for all bitcoin software worldwide to be upgraded and re-programmed.
We propose to start in a very gradual way and make the change
only at the next reward halving which is at 420000 blocks,
which is expected to occur at the end of year 2016.
%and the following one  is  at  630000 blocks (roughly end of 2020).
We also want the new mechanism to share many features of the old system
and we do not propose any revolution, rather an evolution which
keeps all that main premises of the old mechanism.
The only thing we want to achieve is to smooth the reward thresholds to become more continuous and less abrupt.

It is not obvious to see how to design such a mechanism.
We could for example propose to decrease the block reward
to be decremented after every 2016 blocks (roughly every two weeks),
when the $\mbox{\var{difficulty}}$ changes in the current bitcoin software.
This would make it relatively easy to manage for miners.
However the problem is that 210,000 is not a multiple of 2016 and we would like
to be able to compare the new scheme to the old scheme with abrupt changes at every 210,000 blocks.
Therefore we propose to decrement it every 336 blocks.
We have $GCD(210000,2016)=336$, $336=3\cdot 2^4\cdot 7$
and $2016=6\cdot 336$.
This creates a cycle which is still possible to manage for miners and which will always happen at exact boundaries of
two other cycles of 2016 and 210,000 blocks.
Interestingly we have $210,000=336\cdot 5^4$
and the change will happen exactly three times per week as
$3\cdot 336\cdot 10$ minutes is exactly 7 days.
This will make our solution quite elegant
and we obtain one nice closed formula (cf. Fig. \ref{FigureReformProposal} below).

Our new mechanism is designed in order to satisfy the following requirements:

\vskip-7pt
\vskip-7pt
\begin{enumerate}
\item
The block reward should be decremented every 336 blocks,
starting at block 420,336 where it should already be slightly smaller than 25 BTC.

\item
We want to maintain
%the same production of 12.5 BTC per block on average in the interval
%420,000-630,000 and the same rate of monetary base supply at any later period of 4 years,
%in order to to maintain
the exact cap of 21 million BTC ever produced.
\item
This means that at the beginning %for a certain initial fraction of the 4-year period
the reward will be bigger than before.
Then eventually later it will be smaller than before.
\item
We want to have one single continuous curve starting from block 420,000 and one single closed formula.
This turns out to be possible and we can solve our problem with one single mathematical formula.
\item
Currently the reward is halved at each 210,000 blocks.
For blocks up to 209,999 we had 10,5 millions bitcoins produced.
\item
For all blocks up to block 419,999 we will have $10.5+5.25=15.75$ millions of bitcoins produced.
\item
It remains 5.25 million of bitcoins to be produced.
We need to be able to maintain this exact number
with a new reward formula for the period between 420,000 and infinity.
\end{enumerate}
\vskip-5pt

Thus we postulate that  the  following modified
reward mechanism should be introduced at and after block 420000:

\newpage
\vskip-5pt
\vskip-5pt
\begin{figure}
\vskip-0.1cm
\hrule
\vskip0.1cm
\vskip-0.0cm
%\fbox{
%%\includegraphics[bb=0 -20 1200 780,keepaspectratio,scale=0.30,clip]{hitag2l.png}
%}
\begin{enumerate}
\item
At next reward halving block 420,000, the reward is NOT going to change to 12.5 bitcoins.
Instead it will decrease slowly.
\item It will change for {\bf each period of 663 blocks} with small decrements.
\item For any block number $0\leq t< 210,000$ the reward was 50 BTC.
\item For any block number $210,000\leq t< 420,000$ the reward remains 25 BTC.
\item For any block number $t\geq 420,000$ the reward is going to be decreased
following a geometric progression and it should be decremented at each 663 blocks.
More precisely the reward for mining block number $t=336\cdot k$ should be:

%old Satoshi formula starts at 420,000
%sum(210000*12.5/2^k,'k'=0..infinity)/10^6;
\vskip-5pt
\vskip-5pt
$$
\mbox{\var{reward}}_{t\in [336\cdot k\ldots 336\cdot k+335]}
~~=~~
\mathbf{
25.0\cdot (\frac{625}{624})^{1250-k}
}
\mbox{~BTC~~~for any~~} k\geq 1250.
$$

%new
%sum(336*25.0*(625/624)^((-k+1250)),'k'=1250..infinity)/10^6;
\item The actual reward will be rounded to the nearest Satoshi,
the smallest unit in the currency, equal to 1/100,000,000 BTC.

%At the beginning of this period the reward is close to 25 BTC instead of 12.5 BTC before.
\end{enumerate}
\vskip-3pt
\vskip-2pt
\hrule
\vskip-5pt
  \caption{Our proposal for smoothing the miner reward mechanism}
  \label{FigureReformProposal}
\end{figure}
\vskip-5pt

{\bf Remark.}
Yes this means that for the first half on the 4-year period the reward
will be first bigger than before, and only eventually later it will be smaller.
This is inevitable if want to maintain the same production of 21 millions of bitcoins
in the long run and have one single closed formula to use.
%12.5 BTC per block on average in the interval
%420,000-630,000 and a steady continuous pace in the monetary base supply.

{\bf Correctness.}
We give here a calculation sheet in Maple which
proves the correctness of our reward scheme.
For comparison we do it also the previous (original Satoshi) reward scheme.
%% full version maybe %%The original Satoshi mechanism
%% full version maybe %%can be described by the following formula.
%% full version maybe %%Let $t=210000*f$, the reward for any period of time $t\geq 420,000$ is:
%% full version maybe %%$$
%% full version maybe %%\mbox{\var{reward}}_{t\in [210000\cdot f\ldots 210000\cdot f+209999]}
%% full version maybe %%~~=~~
%% full version maybe %%%\mathbf{
%% full version maybe %%12.5\cdot 2^{2-f}
%% full version maybe %%%}
%% full version maybe %%\mbox{~BTC~~~for any~~} f\geq 2.
%% full version maybe %%$$

\vskip-6pt
\vskip-6pt
\begin{verbatim}
>#Satoshi
>15.75+sum(210000*12.5/2^(f-2),'f'=2..infinity)/10^6;
                                          21.00000000
>#New formula
>15.75+sum(336*25.0*(625/624)^((1250-k)),'k'=1250..infinity)/10^6;
                                          21.00000000
\end{verbatim}
\vskip-6pt

{\bf Examples.}
We give here below some concrete examples of rewards with the old formula and the new formula.
We see that with the new scheme for a long time the miner reward is higher than before.
We contend that this property is probably impossible to avoid
if we want to maintain a geometric progression in one single closed formula,
the 21 million cap and continuity at block 420,000.

%block 100410 was first block in year 2011
%block 214563 was first block in year 2013
\vskip-4pt
\vskip-4pt
\begin{figure}[h!]
\vskip-4pt
\vskip-4pt
\hskip-10pt
\hskip-10pt
$$
\begin{array}{|c|c|c|c|c|c|c|c|c|c|}
\hline
\mbox{block}           & 105,000 & 210,000 & 420,000 & 420,336 & 525,000  & 630,000 & 840,000 &1050,000\\
\hline
\mbox{date}            & 01/2011 & 11/2012 & 11/2016 & 11/2016 & 11/2018  & 11/2020 & 11/2024 & 11/2028\\
\hline
\mbox{old formula}      & 50.0       & 25.0 & 12.5    & 12.5    & 12.5     & 6.25    & 3.125    & 1.5625\\
\hline
\mbox{new formula}      & 50.0       & 25.0 & 25.0    & 24.97    & 15.16   & 9.18    & 3.378    & 1.2417\\
\hline
\end{array}
$$
   \caption{Examples of reward with the old and the new system}
     \label{}
\vskip-4pt
\vskip-4pt
\end{figure}
\vskip-4pt

\newpage

\vskip-6pt
\vskip-6pt
\section{Summary and Conclusion}
%\section{\uppercase{Conclusions}}
\label{sec:conclusion}
\vskip-6pt

\noindent
%% full version maybe %%It is sometimes believed
%% full version maybe %%that %collaborative
%% full version maybe %%networks create a lot of value
%% full version maybe %%and their value
%% full version maybe %%grows very substantially as they grow
%% full version maybe %%even though network effects have been
%% full version maybe %%vastly exaggerated, cf. \cite{OdlyzkoMetcalfe}.

In this paper we explain how
bitcoin electronic currency works and
show that the profitability of bitcoin mining
%% full version maybe %%and therefore also the price of bitcoin
depends on a certain cryptographic constant
%which we showed to be at most 1.86.
which we showed to be at most 1.89.
%This was not yet discussed in the open literature
Normally very few people care about this sort
of fine cryptographic engineering details.
However here it is different.
This observation alone allows bitcoin miners
to save
%up to 75,000 USD per day in electricity bills.
%This would be
many millions of dollars each year.
%% full version maybe %%Many other very technical factors also influence
%% full version maybe %%the feasibility of very efficient bitcoin miners
%% full version maybe %%and the price of bitcoin.
The biggest incertitudes however do not come from cryptography.

\vskip-6pt
\vskip-6pt
\subsection{The Unreasonable Incertitudes}
% And Ugly Incentives}
\vskip-5pt

In this paper we argument that investors who build, buy,
or run bitcoin miners are exposed to a number of incertitudes
and risks which strongly depend on the current bitcoin specification
and its future updates.
%We see several major sources of instability of bitcoin.
%in this market.
%% full version maybe %%Bitcoins are unlikely to behave as a real currency,
%% full version maybe %%they do not have legal protection
%% full version maybe %%and they are likely to behave in totally unreasonable ways.
We contend that the lack of stability of bitcoin
goes far beyond the relatively small size of the market compared
to traditional finance.
It can be and to some extent it is
an inherent built-in feature. %which needs to be modified.
In particular
we found the bitcoin 4-year reward halving system very disturbing.
We see absolutely {\bf no reason and no benefit of any kind
to have this sort of mechanism built in a crypto currency.}
On the contrary, we claim that this mechanism is
{\bf unnecessary and harmful}.
It somewhat discredits bitcoin as a stable currency
which could be a reliable store of value.
At the very least it can seriously
limit a wider adoption of bitcoin.
Therefore we propose to change it as soon as possible
and propose a modified reward formula.

%% full version maybe %%Bitcoin have been called digital gold.
%% full version maybe %%However the production of gold does not obey to the same rules.
%% full version maybe %%%Gold cannot be manufactured...
Bitcoin obeys the rules of a new technology venture business with a strong business cycle
and strong incentives for innovation.
Each new generation of bitcoin miners drives the previous generation out of business
each time the production method is changed.
%, and the entry barriers are growing.
In this paper we have shown that one can improve %the 4-th generation of
bitcoin miners also on the algorithmic side and we have achieved a 38 $\%$ improvement.
This is not a lot compared to the fact that mass production of specialized ASIC
circuits have allowed many more substantial improvements simply due
to higher density and highly specialized production.
Will there ever be a fifth generation of miners which brings a more radical change?
We contend that better technology is bound to be invented, would it be quantum bitcoin miners.

More importantly, the specification of the problem to solve is likely to change.
The idea to modify the Proof of Work function in bitcoin have been proposed in May 2013
at Bitcoin conference in San Diego by Dan Kaminsky \cite{KaminskyPredictsEndOfSHA256}.
However this type of change could be flatly rejected by the community
which have heavily invested in ASIC mining with the current technology.

In fact there is a profound reason why such a change is proposed.
According to Dan Kaminsky the production of new bitcoins is now in the hands
of too few people \cite{KaminskyPredictsEndOfSHA256}.
Bitcoin has lost its widely distributed and democratic character.
%It is maybe no longer true that bitcoin belongs to no one, or that it belongs
%to a wider community.
In particular the companies which manufacture ASIC miners
\cite{ButterflyOfficial60WClaimed,KNCMinerspecs,Bitmine.ch} are in the privileged position.
They can decide to whom they sell their hardware
or delay their shipping for no reason and it is profitable for them to do so.
Potentially more competition and wider promotion of
honesty, good reputation and trustworthiness in this business
could solve this problem,
as there is no public authority
which would regulate the bitcoin mining market.

%\newpage

\vfill
\pagebreak

%\vskip44pt
\appendix

\end{document}